\PassOptionsToPackage{switch}{lineno}
\documentclass[twocolumn]{aastex631}

\usepackage[utf8]{inputenc}
\usepackage{graphicx}
\graphicspath{{./}{figures/}}

\usepackage{array}

\usepackage{amsmath}
\usepackage{array}  

\usepackage{enumitem}

\usepackage{hyperref}

\usepackage{booktabs}

\usepackage[mathlines]{lineno}
\received{}
\revised{}
\accepted{}

\submitjournal{\apj}

\shorttitle{\scriptsize Spatial Variations of the $\mathrm{^{12}CO}$ SLED in NGC 7469 from CO(1–0) to CO(8–7) at 120 pc Scales}
\shortauthors{Saravia et al.}

\begin{document}

\title{Spatial Variations of the $\mathrm{^{12}CO}$ Spectral Line Energy Distribution Across the Central Region of NGC~7469 from CO(1–0) to CO(8–7) at 120 pc Resolution}

\correspondingauthor{Alejandro Saravia}
\email{mas6um@virginia.edu}

\author[0000-0003-4546-3810]{A. Saravia}
\altaffiliation{Grote Reber Fellow of the National Radio Astronomy Observatory}
\affiliation{Department of Astronomy, University of Virginia, 530 McCormick Road, Charlottesville, VA 22903, USA}
\affiliation{National Radio Astronomy Observatory, 520 Edgemont Road, Charlottesville, VA 22903, USA}
\affiliation{Universidad Nacional Autónoma de Honduras, Ciudad Universitaria, Tegucigalpa, Honduras}

\author[0000-0003-0057-8892]{L. Barcos-Muñoz}
\affiliation{National Radio Astronomy Observatory, 520 Edgemont Road, Charlottesville, VA 22903, USA}
\affiliation{Department of Astronomy, University of Virginia, 530 McCormick Road, Charlottesville, VA 22903, USA}

\author[0000-0003-2638-1334]{A. S. Evans}
\affiliation{Department of Astronomy, University of Virginia, 530 McCormick Road, Charlottesville, VA 22903, USA}
\affiliation{National Radio Astronomy Observatory, 520 Edgemont Road, Charlottesville, VA 22903, USA}

\author[0000-0001-5434-5942]{P. van der Werf}
\affiliation{Leiden Observatory, Leiden University, P.O. Box 9513, 2300 RA Leiden, The Netherlands}

\author[0000-0001-5429-5762]{K. Harrington}
\affiliation{Joint ALMA Observatory, Alonso de Córdova 3107, Vitacura 763-0355, Santiago, Chile}

\author[0000-0003-0699-6083]{T. Diaz-Santos}
\affiliation{Institute of Astrophysics, Foundation for Research and Technology-Hellas (FORTH), Heraklion, 70013, Greece}
\affiliation{School of Sciences, European University Cyprus, Diogenes Street, Engomi 1516, Nicosia, Cyprus}

\author[0000-0003-4286-4475]{M. S\'{a}nchez-Garc\'{i}a}
\affiliation{Institute of Astrophysics, Foundation for Research and Technology-Hellas (FORTH), Heraklion, 70013, Greece}

\author[0000-0003-3474-1125]{G. C. Privon}
\affiliation{National Radio Astronomy Observatory, 520 Edgemont Road, Charlottesville, VA 22903, USA}
\affiliation{Department of Astronomy, University of Florida, 1772 Stadium Road, Gainesville, FL 32611, USA}
\affiliation{Department of Astronomy, University of Virginia, 530 McCormick Road, Charlottesville, VA 22904, USA}

\author[0000-0002-1000-6081]{S. T. Linden}
\affiliation{Steward Observatory, University of Arizona, 933 N Cherry Avenue, Tucson, AZ 85721, USA}

\author[0000-0003-3498-2973]{L. Armus}
\affiliation{California Institute of Technology, IPAC, 1200 E. California Boulevard, Pasadena, CA 91125, USA}

\author[0000-0002-1912-0024]{V. U}
\affiliation{California Institute of Technology, IPAC, 1200 E. California Boulevard, Pasadena, CA 91125, USA}

\author[0000-0002-6570-9446]{M. Bianchin}
\affiliation{4129 Frederick Reines Hall, Department of Physics and Astronomy, University of California, Irvine, CA 92697, USA}
\affiliation{Instituto de Astrof\' isica de Canarias, Calle V\'ia L\'actea, s/n, E-38205, La Laguna, Tenerife, Spain}
\affiliation{Departamento de Astrof\'isica, Universidad de La Laguna, E-38206, La Laguna, Tenerife, Spain}

\author[0000-0001-7568-6412]{E. Treister}
\affiliation{Instituto de Alta Investigaci\'on, Universidad de Tarapac\'a, Casilla 7D, Arica, Chile}

\author[0000-0002-3139-3041]{Y. Song}
\affiliation{European Southern Observatory, Alonso de Córdova, 3107, Vitacura, Santiago 763-0355, Chile}
\affiliation{Joint ALMA Observatory, Alonso de Córdova, 3107, Vitacura, Santiago 763-0355, Chile}
\affiliation{Max-Planck-Institut für Radioastronomie, Auf dem Hügel 69, 53121 Bonn, Germany}

\author[0009-0005-7950-3288]{C. Riesco}
\affiliation{LUX, Observatoire de Paris, PSL Research University, Sorbonne Université, CNRS, 75014 Paris, France}
\affiliation{Sorbonne Université, CNRS, UMR 7095, Institut d’Astrophysique de Paris, 98bis bd Arago, 75014 Paris, France}

\author[0000-0001-8490-6632]{T. Lai}
\affiliation{California Institute of Technology, IPAC, 1200 E. California Boulevard, Pasadena, CA 91125, USA}

\author[0000-0003-4209-1599]{Y.-H. Teng}
\affiliation{Department of Astronomy, University of Maryland, 4296 Stadium Drive, College Park, MD 20742, USA}

\author[0000-0002-2603-2639]{D. Kakkad}
\affiliation{Centre for Astrophysics Research, University of Hertfordshire, College Lane, Hatfield, AL10 9AB, UK}

\author[0000-0003-3168-5922]{E. Momjian}
\affiliation{National Radio Astronomy Observatory, P.O. Box O, Socorro, NM 87801, USA}

\author[0000-0001-9163-0064]{I. Yoon}
\affiliation{National Radio Astronomy Observatory, 520 Edgemont Road, Charlottesville, VA 22903, USA}

\author[0000-0003-0532-6213]{C.M Lofaro}
\affiliation{Institute of Astrophysics, Foundation for Research and Technology-Hellas (FORTH), Heraklion, 70013, Greece}

\author[0000-0002-5031-4898]{Ming-yi Lin}
\affiliation{University of Toledo, 2801 Bancroft St, Toledo, Ohio, 43606, USA}

\author[0000-0003-4023-8657]{L. Lenkic}
\affiliation{California Institute of Technology, IPAC, 1200 E. California Boulevard, Pasadena, CA 91125, USA}

\author[0000-0003-3638-8943]{N. Torres-Alb\'a}
\affiliation{Department of Astronomy, University of Virginia, 530 McCormick Road, Charlottesville, VA 22903, USA}

\author[0000-0002-1185-2810]{Cosima Eibensteiner}
\altaffiliation{Jansky Fellow of the National Radio Astronomy Observatory}
\affiliation{National Radio Astronomy Observatory, 520 Edgemont Road, Charlottesville, VA 22903, USA}

\author{B. Yeboah}
\affiliation{National Radio Astronomy Observatory, 520 Edgemont Road, Charlottesville, VA 22903, USA}

\author[0009-0002-6248-3688]{J. Nagarajan-Swenson }
\affiliation{Department of Astronomy, University of Virginia, 530 McCormick Road, Charlottesville, VA 22903, USA}
\altaffiliation{Grote Reber Fellow of the National Radio Astronomy Observatory}

\author[0009-0002-2049-9470]{K. Green}
\affiliation{Department of Astronomy, University of Virginia, 530 McCormick Road, Charlottesville, VA 22903, USA}

\author[0000-0002-5828-7660]{S. Aalto}
\affiliation{Department of Space, Earth and Environment, Onsala Space Observatory, Chalmers University of Technology, 43992 Onsala, Sweden}

\author[0000-0002-1568-579X]{D. Kunneriath }
\affiliation{National Radio Astronomy Observatory, 520 Edgemont Road, Charlottesville, VA 22903, USA}

\author[0009-0003-4835-2435]{V. A. Buiten}
\affiliation{Leiden Observatory, Leiden University, P.O. Box 9513, 2300 RA Leiden, The Netherlands}

\author[0000-0002-6650-3757]{J. Kader}
\affiliation{Department of Physics and Astronomy, 4129 Frederick Reines Hall, University of California, Irvine, CA 92697, USA}

\author[0000-0001-7421-2944]{A. M. Medling}
\affiliation{Department of Physics \& Astronomy and Ritter Astrophysical Research Center, University of Toledo, Toledo, OH 43606, USA}

\author[0000-0001-9452-0813]{T. Izumi}
\affiliation{National Astronomical Observatory of Japan, National Institutes of Natural Sciences, Mitaka, Tokyo 181-8588, Japan}

\author[0000-0001-6956-0987]{E. Rodas-Quito}
\affiliation{Universidad Nacional Autónoma de Honduras, Ciudad Universitaria, Tegucigalpa, Honduras}

\author[0009-0006-6594-1516]{G. A. Meza }
\affiliation{Universidad Nacional Autónoma de Honduras, Ciudad Universitaria, Tegucigalpa, Honduras}

\author[0000-0001-6527-6954]{K. L.~Emig}
\affiliation{National Radio Astronomy Observatory, 520 Edgemont Road, Charlottesville, VA 22903, USA}

\author[0000-0001-7291-0087]{J. Surace}
\affiliation{California Institute of Technology, IPAC, 1200 E. California Boulevard, Pasadena, CA 91125, USA}

\author[0000-0002-1233-9998]{D. Sanders}
\affiliation{Institute for Astronomy, University of Hawaii, 2680 Woodlawn Drive, Honolulu, HI 96822, USA}

\begin{abstract}
We present matched-resolution ALMA observations of seven $\mathrm{^{12}CO}$ transitions spanning CO(1--0) to CO(8--7) in NGC~7469, resolving its Seyfert nucleus and circumnuclear star-forming ring at $\sim$120~pc scales. The CO spectral line energy distributions (SLEDs) reveal enhanced nuclear excitation and substantial diversity within the ring. The nucleus exhibits the strongest high-$J$ CO emission relative to CO(1--0), although some ring regions show comparable ratios between the highest observed consecutive transitions. Within the central $\sim1.6$~kpc-diameter field, the circumnuclear ring and surrounding molecular gas contribute $\sim74$--$91\%$ of the recovered CO flux from CO(1--0) through CO(7--6), despite the stronger nuclear excitation. One-component non-LTE modeling of the nucleus and three ring complexes yields molecular gas densities of $n_{\rm H_2}\sim10^3$--$10^5~\mathrm{cm^{-3}}$ and kinetic temperatures from tens to $\sim10^3$~K. The solutions reveal regional diversity rather than a simple progression from nucleus to ring, and the enhanced nuclear CO excitation does not require the highest gas temperature. Illustrative two-component models suggest that a small fraction of hot molecular gas ($T_{\rm kin}\sim800$~K), embedded within a more extended, cooler CO-emitting reservoir, can simultaneously reproduce the observed CO SLED and account for the higher-excitation H$_2$ emission reported in mid-infrared studies. Modified-blackbody modeling of the $461$--$922$~GHz continuum reveals no significant nuclear--ring differences in spectral shape, while the temperature--emissivity degeneracy prevents robust constraints on spatial variations in dust temperature. These findings demonstrate how spatially distinct molecular environments contribute to integrated CO SLEDs, complicating the interpretation of unresolved observations in terms of a single set of physical conditions.
\end{abstract}


\section{Introduction}\label{Intro}

The molecular interstellar medium (ISM) regulates both star formation and the growth of supermassive black holes, making it a fundamental component of galaxy evolution \citep[e.g.,][]{KennicuttEvans2012,Tacconi2020}. Rather than acting as a passive reservoir, molecular gas responds to stellar feedback and active galactic nucleus (AGN) activity through radiative heating, shocks, turbulence, compression, and outflows; processes that can either enhance or suppress subsequent star formation depending on the local physical conditions \citep[e.g.,][]{Hopkins2012,Bolatto_2013}. These conditions are particularly pronounced in nearby luminous and ultraluminous infrared galaxies (LIRGs and ULIRGs), whose large molecular gas reservoirs, intense star formation, and frequent AGN activity provide key laboratories for investigating how feedback shapes the physical and chemical state of the ISM \citep[e.g.,][]{Sanders1996,U_2012}.

Carbon monoxide (CO) serves as the standard tracer of molecular hydrogen (H$_2$) due to its high abundance and bright rotational transitions, while cold H$_2$ is effectively invisible under typical ISM conditions \citep[][]{KennicuttEvans2012}. CO rotational transitions span a wide range of excitation energies and are readily observable at millimeter and submillimeter wavelengths.
The lowest rotational transitions, particularly CO(1--0), trace the bulk of the cold molecular reservoir and provide the basis for estimating molecular gas masses \citep[e.g.,][]{Bolatto_2013}, whereas progressively higher-$J$ transitions probe increasingly excited molecular gas associated with intense star formation, shocks, or radiative heating by an AGN \citep[e.g.,][]{Spaans_2008,vanderWerf_2010,Carilli_2013,Rosenberg_2015}. The CO spectral line energy distribution (CO SLED), constructed from multiple rotational transitions, therefore provides a powerful description of the molecular excitation structure within galaxies. Modeling the CO SLED with non-LTE radiative-transfer calculations can therefore be used to investigate the physical conditions of the molecular gas and how these conditions vary among different galactic environments \citep[e.g.,][]{Weiss2007,Carilli_2013,Pereira-Santaella_2014,Rosenberg_2015,Harrington_2021,Montoya_Arroyave_2024}.

For example, using \textit{Herschel}/SPIRE spectroscopy from CO(4--3) to CO(13--12), \citet{Rosenberg_2015} demonstrated that the shape of the integrated CO SLED provides a sensitive measure of molecular excitation in nearby LIRGs. Their analysis showed that galaxies with relatively strong high-$J$ CO emission cannot generally be reproduced by classical photodissociation-region heating alone, requiring additional energy sources such as X-ray irradiation or mechanical heating. However, integrated CO SLEDs inevitably combine emission from multiple galactic environments within a single beam. Consequently, they characterize the overall excitation of the molecular gas but cannot determine where the different excitation conditions arise within a galaxy or how they relate to distinct physical environments.

Spatially resolved observations at the sensitivity and angular resolution now achievable with ALMA are therefore essential for connecting the excitation components inferred from integrated CO SLEDs to the underlying molecular structures that produce them. While ALMA has enabled sub-arcsecond imaging of molecular gas in nearby galaxies, most resolved studies remain limited to the lower CO transitions (J $\leq$ 3). For example, \citet{Izumi_2020} obtained high-resolution observations of low-$J$ CO and [C\,\textsc{I}] in NGC~7469, revealing clear differences between the AGN-dominated nucleus and the circumnuclear star-forming ring. In contrast, studies extending to higher-$J$ CO transitions generally rely on galaxy-integrated or relatively low-resolution measurements \citep[e.g.,][]{Montoya_Arroyave_2024}. As a result, the spatial distribution and relative contribution of the high-excitation molecular component to the resolved CO SLED remain poorly understood.

To address these limitations, we initiated a pilot program to study spatially resolved ($\sim100$ pc) CO SLEDs in local U/LIRGs using matched-resolution multi-band ALMA observations. This paper presents the first analysis in this series, focusing on NGC~7469, a nearby LIRG at $z=0.01627$ ($D_L\approx70$ Mpc) hosting a Seyfert~1 nucleus and a circumnuclear star-forming ring \citep[e.g.,][]{Genzel_1995,Lai_2022}. Our new high-resolution ALMA datasets include Band~10 observations of this galaxy, providing a resolved CO SLED spanning CO(1--0) through CO(8--7), complemented by observations of the neutral carbon fine-structure line [C\,\textsc{I}]~($^{3}P_{2}\rightarrow{}^{3}P_{1}$). The nearly face-on orientation of NGC~7469 minimizes line-of-sight confusion and enables a clear spatial separation between the AGN and the surrounding circumnuclear star-forming ring.

Recent observations of NGC~7469 with \textit{JWST} have revealed a rich multiphase environment in which the nucleus is dominated by AGN-driven ionization, warm molecular gas, and localized shocks, while the circumnuclear ring contains numerous compact, dust-embedded star-forming regions powered primarily by stellar radiation fields \citep[e.g.,][]{Armus_2023,Bohn_2023,Bianchin_2024}. These observations establish where the principal heating mechanisms operate but do not determine how they contribute to the cold CO-bright molecular reservoir traced by millimeter and submillimeter transitions.

In this paper, we ask how molecular gas excitation differs between the active nucleus and the circumnuclear star-forming ring of NGC~7469, and what physical conditions can account for those differences. We address this question using resolved line ratios and CO SLEDs, combined with a Bayesian non-LTE analysis of the nucleus and three bright star-forming complexes in the ring. We adopt a one-component description as fiducial and evaluate whether the CO measurements warrant an additional component. Complementary modeling of the submillimeter continuum examines the dust spectral properties, while a separate comparison with published \textit{JWST} H$_2$ emission explores molecular conditions beyond those represented by the CO emission alone. By connecting the resolved CO SLEDs to their plausible physical conditions and spatial contributions, this work provides a framework for interpreting galaxy-integrated CO SLEDs and unresolved observations of luminous infrared and high-redshift galaxies.

Throughout this paper we adopt a cosmology with $H_{0}=70~\mathrm{km\,s^{-1}\,Mpc^{-1}}$, $\Omega_{m}=0.30$, and $\Omega_{\Lambda}=0.70$. At the redshift of NGC~7469 ($z=0.01627$), $1''$ corresponds to 330 pc.

\section{DATA}
\label{data}

\subsection{Spectral Line Imaging}
\label{Spectral_lines_imaging}

We present ALMA observations of CO(4--3), CO(6--5), CO(7--6), and CO(8--7), together with [C\,\textsc{I}]~($^{3}P_{2}\rightarrow{}^{3}P_{1}$), complemented by archival CO(1--0), CO(2--1), and CO(3--2) observations \citep{Izumi_2020}. NGC~7469 is a member of the Great Observatories All-sky LIRG Survey (GOALS; \citealt{GOALS_2009}). These datasets sample its nucleus and circumnuclear star-forming ring, covering the central $\sim2$~kpc. Table~\ref{tab:obs_summary} summarizes the original observations, including their native angular and spectral resolutions and largest angular scales. These quantities precede the spatial-frequency selection and common restoring beam adopted for our analysis.

\begin{table*}[tp]
    \centering  
    
    \resizebox{\textwidth}{!}{
    \begin{tabular}{lcccccl}
    \hline
    \hline
    Transition & Rest Frequency (GHz) & ALMA Band & Beam Size (\arcsec) & LAS (\arcsec) & Channel Width (km~s$^{-1}$) & Notes \\
    \hline
    CO(1--0)   & 115.27  & 3  & 0.2/1.6       & 3.3/19.0         & 5.2  & Archival\textsuperscript{a} \\
    CO(2--1)   & 230.54  & 6  & 0.3/1.2       & 4.8/11.1         & 2.4  & Archival\textsuperscript{a} \\
    CO(3--2)   & 345.80  & 7  & 0.3/0.8       & 4.5/7.7          & 1.7  & Archival\textsuperscript{a} \\
    CO(4--3)   & 461.04  & 8  & 0.3/2.5       & 4.3/17.9         & 2.4  & This work\textsuperscript{b} \\
    CO(6--5)   & 691.47  & 9  & 0.3/1.8       & 3.3/12.9         & 3.4  & This work\textsuperscript{b} \\
    CO(7--6)   & 806.65  & 10 & 0.2/1.0       & 2.4/8.6          & 2.9  & This work\textsuperscript{b} \\
    CO(8--7)   & 921.80  & 10 & 0.2/0.9       & 2.5/7.8          & 2.6  & This work\textsuperscript{b} \\
    C\,\textsc{I}(2--1) & 809.34 & 10 & 0.2/1.0 & 2.4/8.6 & 2.9 & This work\textsuperscript{b} \\
    \hline
    \end{tabular}
    }
    \begin{minipage}{\textwidth}
\begin{description}
    \footnotesize
    \item[a] Archival Data: Project ID 2017.1.00078.S, PI: Takuma Izumi.
    \item[b] This work: Project ID 2023.1.00499.S, PI: Loreto Barcos-Mu\~noz.
    \end{description}
\end{minipage}
    
    \caption{Summary of molecular line observations for NGC~7469 used in this work. For reference, 1\arcsec{} corresponds to 330~pc at $z = 0.01627$. The first and second columns list the molecular transition and its rest frequency. The third column indicates the corresponding ALMA band. The fourth and fifth columns give the synthesized beam size (12 m / 7 m arrays) and the largest angular scale (LAS) recovered, respectively. When two values are listed, they correspond to the 12 m and 7 m arrays, respectively, combined using \texttt{concat} in CASA. The sixth column reports the native spectral resolution in km~s$^{-1}$, and the final column indicates the project ID associated with the original dataset. The CO(5--4) transition cannot be observed from ground-based facilities, as the atmospheric transmission at its frequency is effectively opaque.}
    \label{tab:obs_summary}
\end{table*}

The measurement sets were reprocessed using the \texttt{ScriptForPI} scripts supplied by the ALMA pipeline \citep{alma_pipeline}, with the corresponding CASA versions. We inspected the calibration products and associated weblogs. For the CO(4--3), CO(6--5), and CO(7--6) 12~m data, we used the pipeline continuum-subtracted measurement sets. For CO(8--7), we identified line-free channels and subtracted the continuum using \texttt{uvcontsub}.

Where available, the 12~m and 7~m datasets were combined using \texttt{concat} and jointly imaged. We tested different relative visibility weights to assess the contributions of the two arrays to the reconstruction. We also inspected the residual maps and point-spread functions (psf) of the final images for coherent residual structures and to ensure the psf is gaussian-like.

Imaging was performed with \texttt{tclean} in CASA \citep{CASA, CASA_2022}, using the \texttt{standard} gridder, \texttt{multiscale} deconvolution, and \texttt{briggsbwtaper} weighting with a robust parameter of $0.5$. All transitions were imaged with $0.04\arcsec$ pixels and restored with a circular $0.32\arcsec$ beam. We retained projected baselines longer than $30\,\mathrm{k}\lambda$ with the \texttt{uvrange} parameter in CASA, imposing a common minimum spatial frequency while retaining dataset-dependent maximum baselines. No additional Gaussian $uv$ taper was applied.

We used multiscale components of $0$, $8$, $24$, and $72$ pixels and generated cleaning masks with \texttt{auto-multithresh} \citep{casa_multi_thres}. The cleaning threshold was set to twice the rms estimated from line-free channels in an initial undeconvolved cube, with a maximum of $50{,}000$ iterations. Cleaning was restricted to primary-beam responses above $0.2$, and the final cubes were corrected for primary-beam attenuation.

We characterize the image noise using the median spatial standard deviation across selected line-free channels within a common circular region of radius $2\arcsec$. The resulting rms values are $0.43$, $0.15$, $0.96$, $2.73$, $7.33$, $25.93$, and $26.57$~mJy~beam$^{-1}$ for CO(1--0), CO(2--1), CO(3--2), CO(4--3), CO(6--5), CO(7--6), and CO(8--7), respectively. The corresponding channel widths are $20$, $20$, $10$, $21$, $10$, $10$, and $10$~km~s$^{-1}$.

Within the same region, the maximum channel intensities over radio velocities of $4650$- $5050$~km~s$^{-1}$ correspond to peak-channel S/N values of approximately $47$, $506$, $196$, $118$, $59$, $16$, and $11$, respectively, using the rms estimates above. CO(8--7) therefore has the lowest peak-channel S/N among the observed CO transitions. Its relatively weak emission and the small FOV, which makes the detection of the ring regions particularly more challenging due to the increase noise toward the edges.
 
\begin{figure*}[tp]
    \centering
    \includegraphics[width=1\textwidth]{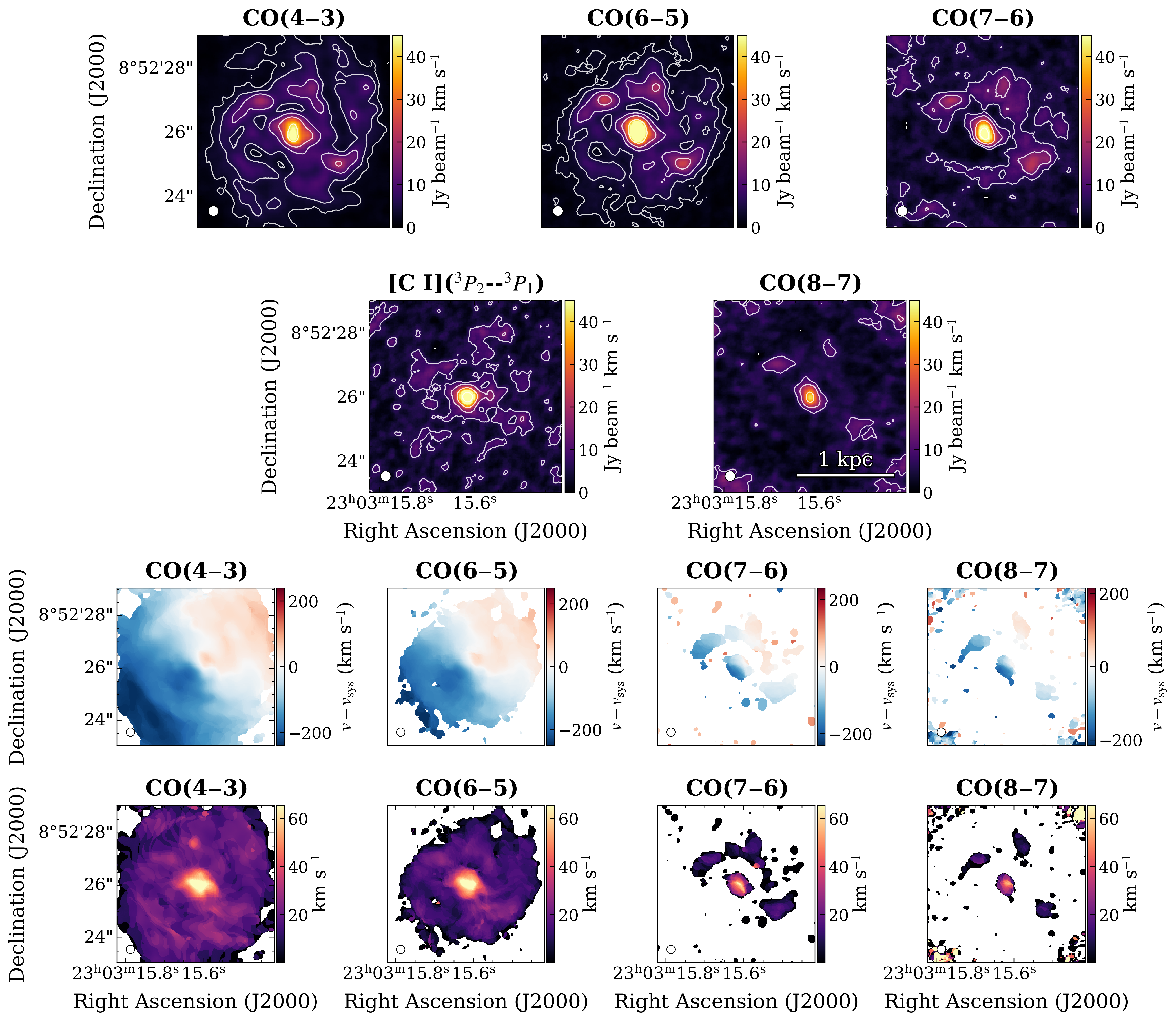}
    \caption{
    ALMA moment maps of CO(4--3), CO(6--5), CO(7--6),CO(8--7), and [C\,\textsc{I}]~($^{3}P_{2}\rightarrow{}^{3}P_{1}$) in the central region of NGC~7469, at a common angular
    resolution of $0.32\arcsec$. The top two rows show integrated intensities (moment~0). The third and fourth rows show the intensity-weighted velocity fields (moment~1) and velocity dispersions (moment~2), respectively, for the four CO transitions. Velocities are relative to the adopted systemic velocity, $v_{\rm sys}=4925~\mathrm{km\,s^{-1}}$. Positive moment-zero contours are drawn at $5$, $10$, $20$, $40$, $80$, and $160$ times the estimated rms of each map, where these levels lie below the peak; dashed contours mark $-3$ times the rms where present. Color bars indicate integrated intensity ($\mathrm{Jy\,beam^{-1}\,km\,s^{-1}}$), velocity offset, and velocity dispersion ($\mathrm{km\,s^{-1}}$). The synthesized beam is shown in the lower-left corner of each panel. The CO(8--7) moment-zero panel includes a 1~kpc scale bar, adopting a distance of 70~Mpc.}
    \label{fig:CO_moment0_panel}
\end{figure*}

The matched-resolution cubes were used for the moment maps in Figure~\ref{fig:CO_moment0_panel}, the resolved line-ratio and CO SLED measurements, and the aperture analysis. Integrated-intensity, intensity-weighted velocity, and velocity-dispersion maps were generated using \texttt{immoments}, with visual inspection of the adopted emission masks. The limitations affecting extended-emission recovery are discussed in Section~\ref{sec:totalfluxes}.

\subsection{Continuum Imaging}
\label{Continuum_Imaging}

\begin{figure*}[tp]
    \centering
    \includegraphics[width=\textwidth]{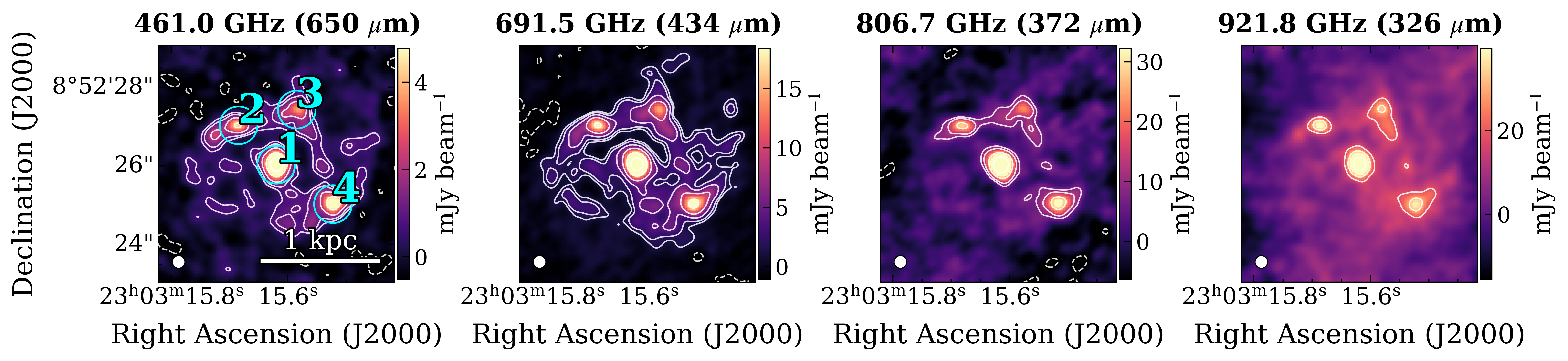}
    \caption{
    ALMA continuum maps of the central region of NGC~7469 at approximately 461, 691, 807, and 922~GHz, restored to a common $0.32\arcsec$ beam. Color bars give flux density in $\mathrm{mJy\,beam^{-1}}$; the color ranges are adjusted independently to emphasize the morphology in each panel. Solid white contours are drawn at $3$, $5$, $10$, $20$, $40$, $80$, and $160$ times the estimated rms of each map, where these levels lie below the peak. Dashed contours mark $-3$ times the rms where present.
    The synthesized beam is shown in the lower-left corner of each panel. Cyan circles in the first panel mark the four $0.96\arcsec$-diameter apertures R1--R4, defined in Section~\ref{grid}. The 1~kpc scale bar adopts a distance of 70~Mpc. }
    \label{fig:Continuum}
\end{figure*}

We constructed continuum images at approximately $461$, $691$, $807$, and $922$~GHz using line-free channels from the original measurement sets (Figure~\ref{fig:Continuum}). We inspected the measurement sets and flagged channels containing line emission before continuum imaging. Where available, the 12~m and 7~m datasets were combined using \texttt{concat}, following the array-combination procedure described in Section~\ref{Spectral_lines_imaging}.

Continuum imaging was performed using \texttt{tclean} with the \texttt{standard} gridder and \texttt{multiscale} deconvolver, employing \texttt{briggs} weighting with a robust parameter of $0.5$. All images were restored with a circular $0.32\arcsec$ beam, providing a common angular resolution across the four frequencies.

The rms estimates used to define the continuum-map contours are $0.21$, $0.43$, $2.54$, and $6.20$~mJy~beam$^{-1}$ at $461$, $691$, $807$, and $922$~GHz, respectively. These estimates were obtained by iterative $3\sigma$ clipping of the lower-intensity $70\%$ of finite pixels in each continuum image.

\section{Methodology}

\subsection{Spatial sampling and aperture definition}
\label{grid}

We employ two complementary spatial-sampling approaches throughout this work. First, we use a beam-sized hexagonal grid to characterize the spatial distribution of CO emission and excitation across the central region of NGC~7469, with detected coverage varying among transitions. Second, we extract spectra and integrated line intensities within four circular apertures centered on the nucleus and three bright regions of the circumnuclear star-forming ring, selected to have detections in all seven observed CO transitions. These apertures are used for the non-LTE excitation analysis and the continuum study presented below.

For the spatially resolved analysis, we follow the grid-based methodology of \citet{Saravia_2025}. Each grid element has a hexagonal shape with an area approximately equal to that of the common $0.32\arcsec$ circular Gaussian restoring beam, corresponding to $0.116~\mathrm{arcsec}^{2}$. The grid therefore samples the emission on approximately the resolution scale, while retaining the spatial correlations imposed by the beam.

For each transition, we identify candidate emission intervals in the spectrum extracted from each hexagon using the rms of the noise measured in line-free channels $\sigma_{\rm ch}$. The automated selection requires emission above $3\sigma_{\rm ch}$ over at least three to five consecutive channels, depending on the transition's channel width. We visually inspect all profiles, including those classified as nondetections, to validate the candidate intervals, identify line emission missed by the automated channel-count criterion, and reject noise fluctuations or imaging artifacts incorrectly identified as emission. The reviewed emission intervals define the integration windows and detection flags used in the subsequent analysis.

Integrated line measurements are calculated by summing the original spectrum over the accepted emission intervals and multiplying by the transition's channel width. Absorption-like features were excluded from the emission integration, and no correction for possible absorption was applied. The conditional thermal uncertainty is $\sigma_{\rm ch}\delta v\sqrt{N_{\rm ch}}$, where $\delta v$ is the channel width and $N_{\rm ch}$ is the total number of unique integrated channels. This estimate assumes independent channels and fixed integration intervals. We separately adopt absolute flux-calibration uncertainties of 10\% for CO(1--0), CO(2--1), CO(3--2), and CO(4--3), and 20\% for CO(6--5), CO(7--6), and CO(8--7), following the nominal ALMA band-dependent accuracies but adopting a conservative 10\% for CO(1--0).

The resulting measurements are used to construct the line-ratio maps, CO SLEDs, and excitation diagnostics presented in Sections~\ref{sec:ratios} and~\ref{sec:sleds}. Ratio maps use velocity-integrated brightness temperatures, $W_J\equiv\int T_{{\rm b},J}\,dv$, and include only hexagons detected in both transitions with positive, finite integrated measurements. We denote their ratios by $R_{ab}=W_{a\rightarrow a-1}/W_{b\rightarrow b-1}$. Reported map medians are medians of the individual hexagon ratios, not ratios of summed intensities. Brightness-temperature ratios are distinguished from flux-normalized SLEDs: the $J^2$ reference applies to the latter for optically thick, thermalized emission in the Rayleigh--Jeans limit and a common emitting area. 

For the aperture analysis, we define four circular regions, R1--R4, each with a diameter of $0.96\arcsec$ ($\sim320$~pc), centered on the nucleus and three prominent emission complexes in the circumnuclear star-forming ring. R1 encompasses the nucleus, while R2--R4 sample the ring. The apertures are overlaid on the continuum images in Figure~\ref{fig:Continuum}, and the extracted flux densities are listed in Table~\ref{tab:continuum_fluxes}. These apertures are adopted consistently for the continuum analysis and CO SLED modeling described in Sections~\ref{sec:continuum_method} and~\ref{sec:radex_method}. Their diameter is distinct from the angular resolution of the images, and the four selected regions are treated as case studies rather than a statistically representative sample of the molecular disk.

\subsection{Continuum aperture photometry and modified-blackbody modeling}
\label{sec:continuum_method}

We measure continuum flux densities within the four circular apertures defined in Section~\ref{grid}, using the continuum images at approximately 461, 691, 807, and 922~GHz, corresponding to wavelengths of approximately 650, 434, 372, and 326~$\mu$m, respectively.

We model the aperture-integrated continuum emission using an optically thin, single-temperature modified blackbody,
\begin{equation}
    S_{\nu}
    =A\left(\frac{\nu}{\nu_{0}}\right)^{\beta}
    B_{\nu}(T_{\rm dust}),
    \label{eq:mbb_model}
\end{equation}
where $A$ is a normalization constant, $\nu_{0}=691$~GHz is the reference frequency, $\beta$ is the dust emissivity index, and $B_{\nu}(T_{\rm dust})$ is the Planck function evaluated at dust temperature $T_{\rm dust}$. Optical thinness and a single temperature are assumptions of this spectral-shape model. Because the normalization absorbs the solid-angle and dust-mass dependent terms, the fits characterize the continuum shape rather than provide a dust-mass estimate.

The fiducial fits use all four continuum measurements at approximately 461, 691, 807, and 922~GHz. The adopted uncertainty for each measurement combines the aperture-photometry and absolute flux-calibration terms in quadrature. The fit assumes independent errors between the measurements; calibration covariance is not modeled.

We evaluate the model over a fixed temperature grid spanning $20\leq T_{\rm dust}\leq100$~K, chosen to encompass the characteristic far-infrared dust temperatures inferred for nearby LIRGs \citep[e.g.,][]{Herrero-Illana_2019}. At each fixed temperature, the normalization $A$ and emissivity index $\beta$ are optimized by weighted nonlinear least squares, with $-2\leq\beta\leq5$. The fit statistic is
\begin{equation}
    \chi^{2}(T_{\rm dust})
    =\sum_i
    \frac{\left[S_{\nu_i}^{\rm obs}-S_{\nu_i}^{\rm mod}\right]^2}
         {\sigma_{S_{\nu_i}}^2},
    \label{eq:mbb_chi2}
\end{equation}
where the sum extends over the four continuum measurements included in the fiducial fit. The minimum of $\chi^{2}(T_{\rm dust})$ defines the nominal best fitting model. This curve is a profile fit statistic, obtained by optimizing the other parameters at each temperature, rather than a marginalized temperature posterior.

Because the available continuum measurements allow a broad covariance between $T_{\rm dust}$ and $\beta$, we emphasize the continuum spectral shape and the family of acceptable $(T_{\rm dust},\beta)$ combinations. The nominal best-fitting parameters are not interpreted as unique determinations of the intrinsic dust properties. Temperatures obtained by fixing $\beta$ are conditional estimates and are not imposed as constraints on the gas kinetic temperature. Indeed, joint line and continuum modeling of luminous star-forming galaxies has revealed $T_{\rm kin} T_{\rm dust}>2$ \citep{Harrington_2021,Harrington2025}. A joint treatment of the CO and dust SEDs, allowing for distinct gas and dust temperatures, is deferred to future work. The results are presented in Section~\ref{sec:dust_mbb}.

\subsection{Non-LTE excitation modeling}
\label{sec:radex_method}

The CO SLEDs extracted from the four representative apertures are modeled using a non-LTE radiative-transfer analysis based on RADEX \citep{RADEX} through the \texttt{SpectralRadex} interface. We adopt a one-component model as the fiducial description and evaluate an additive two-component model to test whether the additional parameters are warranted. A component denotes an effective set of excitation conditions within an aperture; a successful one-component fit does not establish that the unresolved molecular medium consists of a single physical phase.

For each component, the excitation calculation depends on kinetic temperature $T_{\rm kin}$, molecular hydrogen density $n_{\rm H_2}$, and CO column density per unit velocity width $N_{\rm CO}/\Delta v$. The adopted linewidths are based on Gaussian fits to the aperture-integrated CO profiles, using two Gaussians for the nucleus and one for each ring aperture, as motivated by visual inspection of the line shapes. The fixed model settings and adopted linewidths are listed in Table~\ref{tab:radex_grid} in Appendix~\ref{appendix:radex}.

The predicted integrated brightness temperatures are scaled by an effective filling factor, $\phi$, such that $W_i^{\rm mod}=\phi W_i^{\rm RADEX}$. This factor accounts for dilution of the intrinsic model brightness in the aperture measurement and is not independently converted into a source size. The current aperture analysis includes all seven observed CO transitions, from CO(1--0) through CO(8--7).

Each model is compared with the observed integrated brightness temperatures using
\begin{equation}
    \chi^2
    =\sum_i
      \frac{\left[W_i^{\rm obs}-W_i^{\rm mod}\right]^2}
           {\sigma_i^2},
    \label{eq:radex_chi2_method}
\end{equation}
where $\sigma_i$ denotes the supplied aperture uncertainty, including measurement and calibration terms. We assume independent Gaussian errors, giving a likelihood proportional to $\exp(-\chi^2/2)$; calibration covariance between transitions is not included. The physical-parameter priors are uniform in logarithm over $10\leq T_{\rm kin}\leq2000$~K, $10^1\leq n_{\rm H_2}\leq10^7~\mathrm{cm^{-3}}$, and $10^{15}\leq N_{\rm CO}/\Delta v\leq10^{21}~\mathrm{cm^{-2}\,(km\,s^{-1})^{-1}}$. We adopt a uniform prior on $0\leq\phi\leq1$. For the fiducial model, the likelihood is analytically marginalized over $\phi$, and the physical-parameter posterior is integrated using grid-cell weights consistent with the stated priors.

We summarize the marginalized posterior distributions by their medians and 16th--84th percentile credible intervals. Derived quantities, including the molecular thermal-pressure $P_{\rm H_2}/k_{\rm B}\equiv n_{\rm H_2}T_{\rm kin}$, are calculated from joint posterior samples to account for parameter correlations. The marginal medians therefore summarize individual parameters and need not together describe a representative model. We assess numerical accuracy through comparisons between refined grids and independent direct RADEX calculations, distinguishing numerical errors from the parameter degeneracies and sensitivity to the adopted prior bounds.

The two-component model combines the component intensities through independent filling factors,
\begin{equation}
    W_i^{\rm mod}
    =\phi_1W_{i,1}^{\rm RADEX}+\phi_2W_{i,2}^{\rm RADEX}.
    \label{eq:two_component_model_method}
\end{equation}
Both components use the same physical-parameter priors as the fiducial model. Their emission is added without foreground absorption between components, and the independent filling factors are not constrained to sum to unity. We order the fitted pair by temperature, $T_1<T_2$, solely to remove label ambiguity. The lower- and higher-temperature labels do not imply a corresponding ordering in density or excitation, or establish distinct cold and warm phases. Model comparison uses the Bayesian evidence, which integrates the likelihood over the adopted parameter priors, rather than the improvement in minimum $\chi^2$ alone. The numerical implementation, grid refinement, validation, and evidence calculations are described in Appendix~\ref{appendix:radex}.

\subsection{Supplementary excitation tests}
\label{sec:h2_support}

We perform two exploratory tests to place the fiducial CO modeling in the context of independent molecular diagnostics.

First, we compare the nuclear CO SLED with predictions for selected physical conditions motivated by \citet{Izumi_2020}, whose analysis combined lower-$J$ CO, $^{13}$CO, and [C,\textsc{I}] emission. This test examines whether those selected conditions can also account for the upper CO transitions observed here. It does not reproduce their full multi-tracer analysis or test all conditions compatible with their results. The parameter choices, extraction conventions, and comparison statistics are given in Appendix~\ref{app:izumi_comparison}.

Second, we examine whether the CO-based descriptions account for the mid-infrared H$_2$ emission, using the extinction-corrected S(1)--S(7) fluxes and quoted errors published by \citet{Lai_2022}. We extract separate CO measurements in their nominal $0.8\arcsec$ apertures, with LR7, LR9, and LR1 corresponding to our R2--R4; the prefix LR distinguishes their region identifiers from ours. These measurements are distinct from the fiducial $0.96\arcsec$ CO-only fits. From the published H$_2$ fluxes, we calculate apparent rotational temperatures and compare the emission with predictions from the CO-based descriptions and exploratory two-component examples. The predictions use optically thin, collisionally excited populations from \citet{Neufeld_2026}, allowing the emitting ortho-to-para ratio (OPR) to vary. The H$_2$ fluxes are literature measurements; the excitation diagnostics and model predictions are calculated here. Appendix~\ref{app:h2_comparison} describes the assumptions, aperture limitations, and error tests, while Appendix~\ref{app:joint_examples} presents the illustrative optimized solutions and their diagnostics. These examples are not joint posterior constraints.

\section{Results}
\label{results}

\subsection{CO Morphology and Kinematics}
\label{sec:morphology}

Figure~\ref{fig:CO_moment0_panel} presents the matched-resolution integrated-intensity (moment-0) maps of CO(4--3), CO(6--5), CO(7--6), CO(8--7), and [C\,\textsc{I}]~($^{3}P_{2}\rightarrow{}^{3}P_{1}$). We also present the intensity-weighted velocity (moment-1) and velocity-dispersion (moment-2) maps for the four CO transitions. For comparison, images of CO(1--0) through CO(3--2) at comparable angular resolution are presented by \citet{Izumi_2020}. The detected lower-$J$ emission extends to larger projected radii, tracing a more extended molecular distribution than is detected in the higher-$J$ maps presented here.

The moment-zero maps reveal a transition-dependent distribution of the detected CO emission. CO(4--3) traces the nucleus, the circumnuclear ring, and emission connecting the brighter molecular complexes. The higher-$J$ emission is increasingly concentrated toward the nucleus and localized regions within the ring. CO(7--6) and CO(8--7) therefore occupy a smaller detected portion of the molecular structure traced by CO(4--3). The common minimum $uv$ distance can filter extended emission, while differences in sensitivity also affect the detected extent of each transition. These maps therefore describe the emission recovered on the sampled angular scales rather than establish an intrinsic outer boundary to the higher-$J$ emitting gas.

The moment-one maps show a coherent velocity gradient across the nucleus and circumnuclear ring. Where the transitions are jointly detected, the higher-$J$ emission follows the same redshifted and blueshifted structures traced by CO(4--3) and CO(6--5), consistent with a common large-scale rotation pattern. The moment-two maps show a pronounced central enhancement, with broader emission profiles toward the nucleus than across most of the ring. This enhancement can include contributions from unresolved velocity gradients and overlapping spectral components, in addition to turbulent motions.

\subsection{Total fluxes}
\label{sec:totalfluxes}


The recovery of extended emission is limited by both the original observations and the spatial-frequency selection adopted for the matched analysis. The datasets have different sensitivities and baseline distributions, leading to transition-dependent sensitivity to faint and extended structures. Applying a common minimum $uv$ distance removes shorter baselines and can further reduce sensitivity to extended emission. Because the maximum baselines remain dataset dependent, this selection does not match the complete spatial-frequency coverage. A common minimum $uv$ distance and restoring beam do not guarantee identical spatial responses or equal recovered-flux fractions across transitions.

Table~\ref{tab:total_fluxes} lists the recovered ALMA line fluxes, the number of detected hexagons in each transition, and the corresponding Herschel measurements \citep{Rosenberg_2015, Kamenetzky_2016}. However, the differences cannot be attributed solely to interferometric spatial filtering or to the common $uv$ selection, because the measurements also differ in aperture coverage and sensitivity. Absorption-like features are also present in the CO(7--6) and CO(8--7) line profiles. These features were excluded from the emission integration, and no correction for possible absorption was applied. If they represent genuine absorption, the observed profiles may underestimate the underlying emission, contributing to the difference between the recovered ALMA fluxes and the Herschel measurements.

We consequently interpret the resolved maps and SLEDs as characterizing the recovered emission in individual regions over the retained angular scales, rather than the complete molecular emission of the galaxy.

\begin{table}
\centering
\caption{Number of detected hexagons and comparison of ALMA
and Herschel CO line fluxes.}
\label{tab:total_fluxes}
\begin{tabular}{lccc}
\hline\hline
Transition & $N_{\rm hex}$ & ALMA & Herschel \\
 & & \multicolumn{2}{c}{($10^{-17}$ W m$^{-2}$)} \\
\hline
CO(4--3) & 320 & $2.13 \pm 0.21$ & $2.69 \pm 0.43$ \\
CO(6--5) & 172 & $2.43 \pm 0.49$ & $4.34 \pm 0.69$ \\
CO(7--6) &  71 & $2.08 \pm 0.44$ & $3.45 \pm 0.55$ \\
CO(8--7) &  39 & $0.34 \pm 0.25$ & $2.91 \pm 0.47$ \\
\hline
\end{tabular}
\tablecomments{
$N_{\rm hex}$ is the number of detected hexagons in each transition,
before imposing joint detections with other lines.
}
\end{table}

\subsection{CO Line-Ratio Maps}
\label{sec:ratios}

\begin{figure*}[tp]
    \centering
    \includegraphics[width=1\textwidth]{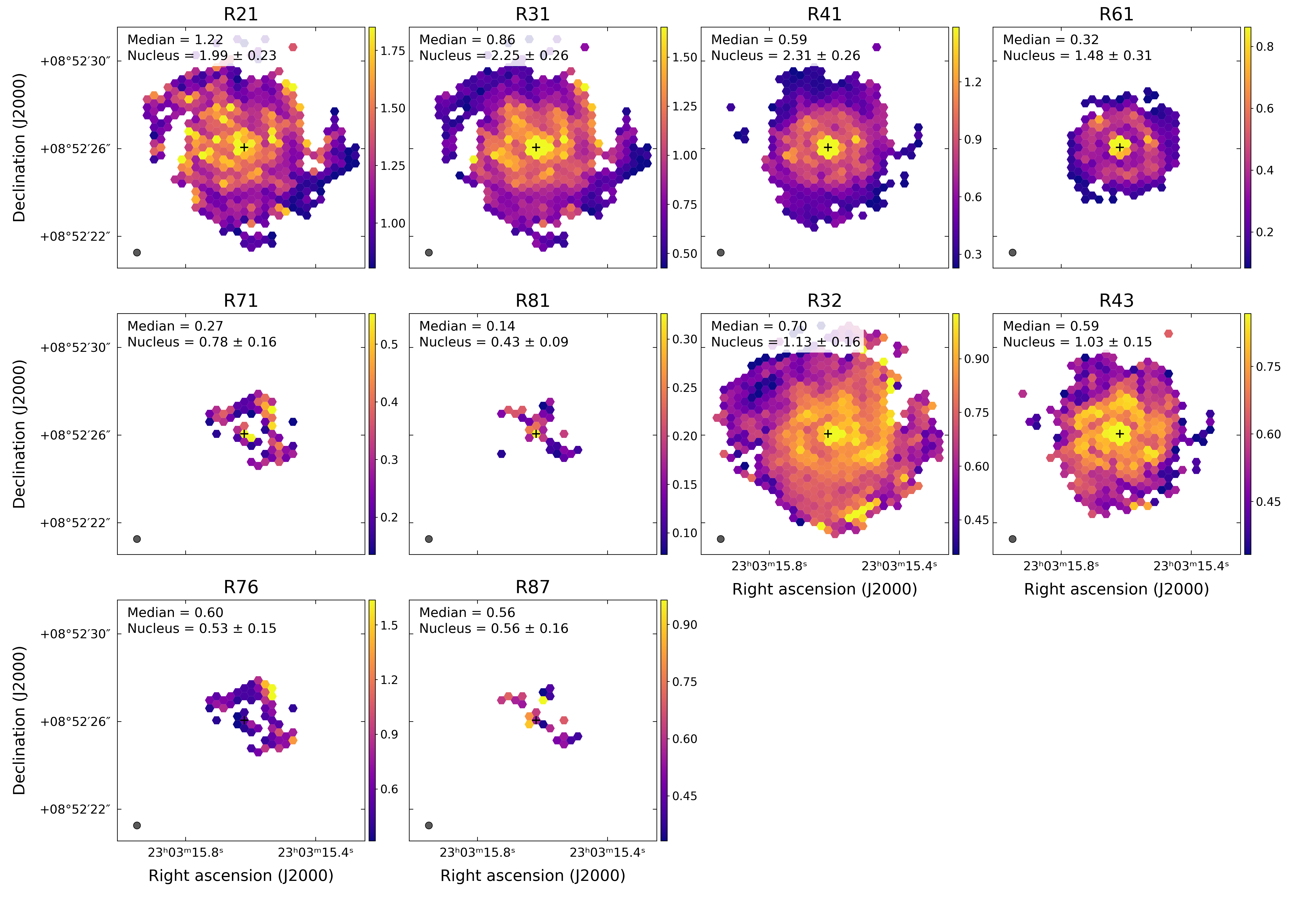}
    \caption{CO line ratio maps of the central region of NGC~7469. The first six panels show ratios normalized to CO(1--0), $R_{J1}$, while the remaining four panels show ratios between consecutive observed transitions, $R_{J,J-1}$. Because CO(5--4) is not accessible from the ground, no consecutive ratio is shown between CO(4--3) and CO(6--5). Colors indicate the integrated brightness-temperature ratio in each beam-sized hexagonal element. The black cross marks the nucleus, and the synthesized beam is shown in the lower-left corner of each panel. The median ratio over all detected hexagons and the value measured at the nucleus are listed in the upper-left corner.}
    \label{fig:ratios}
\end{figure*}

Figure~\ref{fig:ratios} presents the CO line-ratio maps at the common angular resolution. We consider the ratios normalized by CO(1--0), $R_{21}$, $R_{31}$, $R_{41}$, $R_{61}$, $R_{71}$, and $R_{81}$, together with the consecutive-transition ratios $R_{32}$, $R_{43}$, $R_{76}$, and $R_{87}$. These ratios use velocity-integrated brightness temperatures, measured over a common velocity interval across all transitions in each hexagon. Optically thick, thermalized emission in the Rayleigh--Jeans limit, with common emitting areas and velocity profiles, provides a reference ratio near unity.

The ratios relative to CO(1--0) show a pronounced nuclear enhancement. Their detected-sample medians are approximately $1.22$, $0.86$, $0.59$, $0.32$, $0.27$, and $0.14$ for $R_{21}$ through $R_{81}$, respectively. In the central nuclear hexagon, the corresponding values are $1.99\pm0.23$, $2.25\pm0.26$, $2.31\pm0.26$, $1.48\pm0.31$, $0.78\pm0.16$, and $0.43\pm0.09$. The quoted uncertainties represent the propagated statistical and adopted flux-calibration uncertainties. The nuclear contrast is particularly strong in $R_{41}$ and $R_{61}$, whose nuclear values exceed their respective sample medians by factors of approximately $3.9$ and $4.7$.

The spatial coverage differs substantially among these ratios. After applying the joint-detection, spectral-coverage, and integrated signal-to-noise requirements, the six CO(1--0)-normalized maps contain $414$, $413$, $304$, $170$, $53$, and $29$ hexagons, respectively. Their medians therefore describe different selected populations, with the highest-$J$ ratios sampling primarily the nucleus and bright circumnuclear complexes. A sequence constructed from these medians would therefore summarize transition-dependent detected samples rather than the CO SLED of a fixed spatial population. Differences in sensitivity and spatial filtering also affect the detected extent of each map.

The lower-$J$ ratio maps show spatial structure beyond the nuclear enhancement, without delineating the circumnuclear ring as a uniformly elevated feature. The $R_{41}$ and $R_{61}$ maps show a strong central enhancement surrounded by spatially varying ratios. The more sparsely sampled $R_{71}$ and $R_{81}$ maps retain detections in localized circumnuclear complexes as well as the nucleus. The detected emission therefore does not follow a smoothly declining radial sequence, although the changing sensitivity and spatial coverage prevent interpreting undetected positions as an absence of higher-$J$ emission.

The consecutive-transition ratios clarify which part of the SLED drives the nuclear contrast. The detected-sample median $R_{32}$ is $0.70$, compared with $1.13\pm0.16$ in the central hexagon, while $R_{43}$ increases from $0.59$ to $1.03\pm0.15$. At the upper end of the observed SLED, however, the nuclear ratios are close to their detected-sample medians: $R_{76}=0.53\pm0.15$ in the nucleus compared with $0.60$ over the selected sample, and $R_{87}=0.56\pm0.16$ compared with $0.56$. Thus, the nucleus is strongly enhanced when the upper CO lines are compared with CO(1--0), but is less exceptional in the relative strengths of the highest consecutive transitions. Bright regions within the ring can exhibit comparable upper-SLED ratios even when their ratios relative to CO(1--0) are lower.

These maps show that the excitation pattern cannot be described by a radial gradient alone. The strongest overall enhancement occurs toward the nucleus, while localized regions within the ring retain substantial emission in the highest observed transitions. The resolved SLEDs and excitation index, discussed in Section~\ref{sec:sleds}, provide complementary summaries of this spatial variation.

\subsection{Resolved CO Spectral Line Energy Distributions and Excitation Index}
\label{sec:sleds}

\begin{figure*}[tp]
\centering

\includegraphics[width=0.51\textwidth]{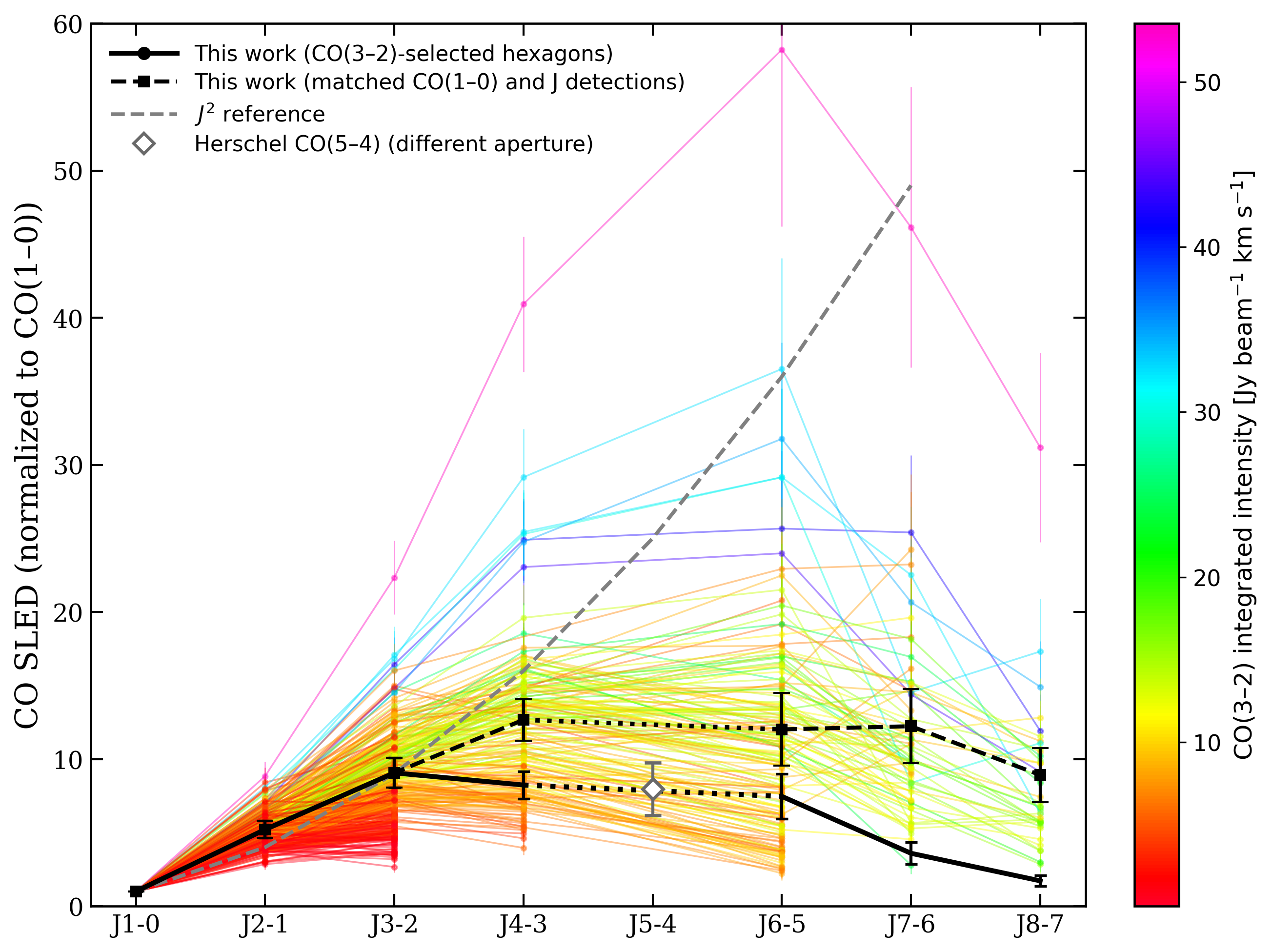}
\hfill
\includegraphics[width=0.48\textwidth]{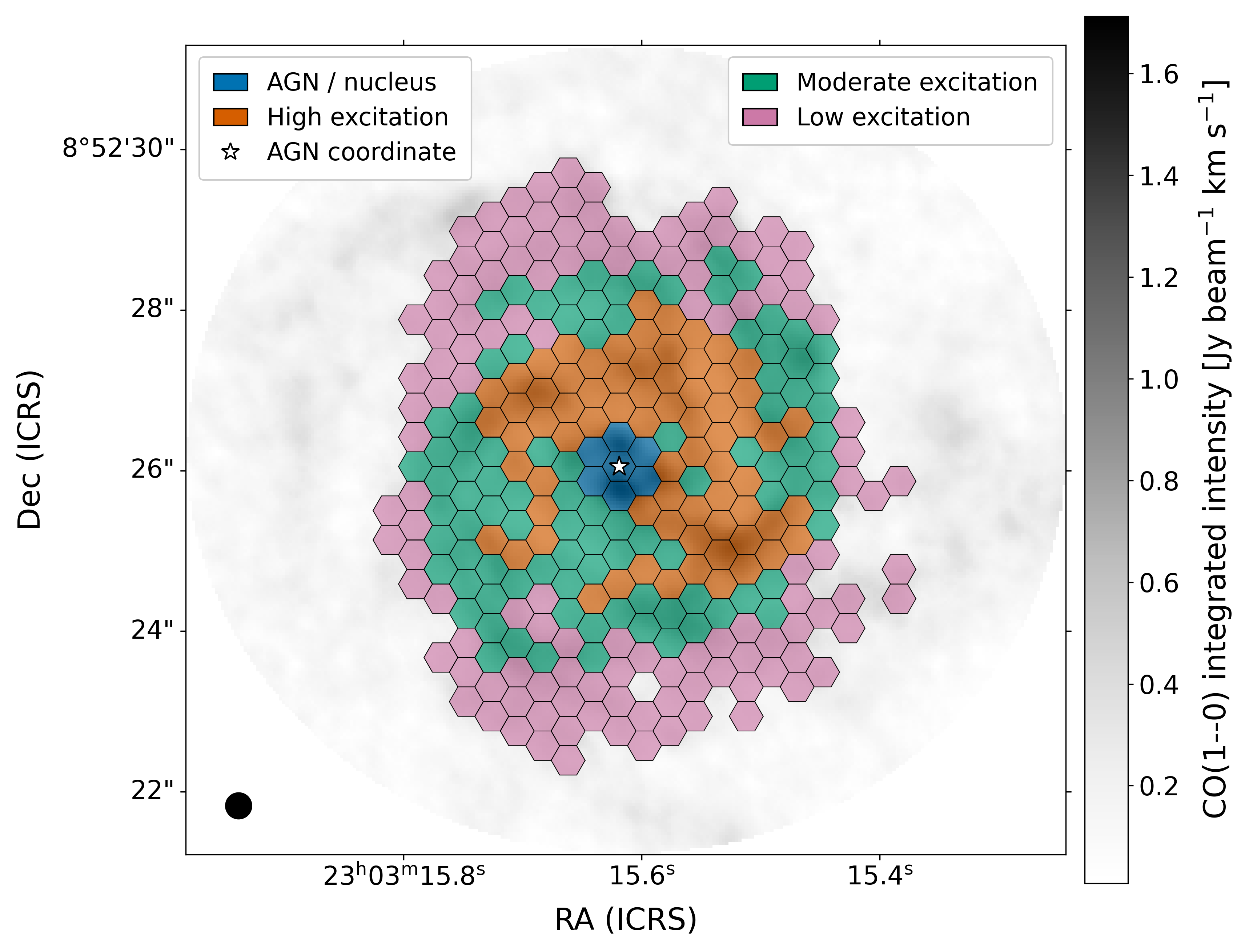}

\vspace{0.3cm}

\includegraphics[width=0.45\textwidth]{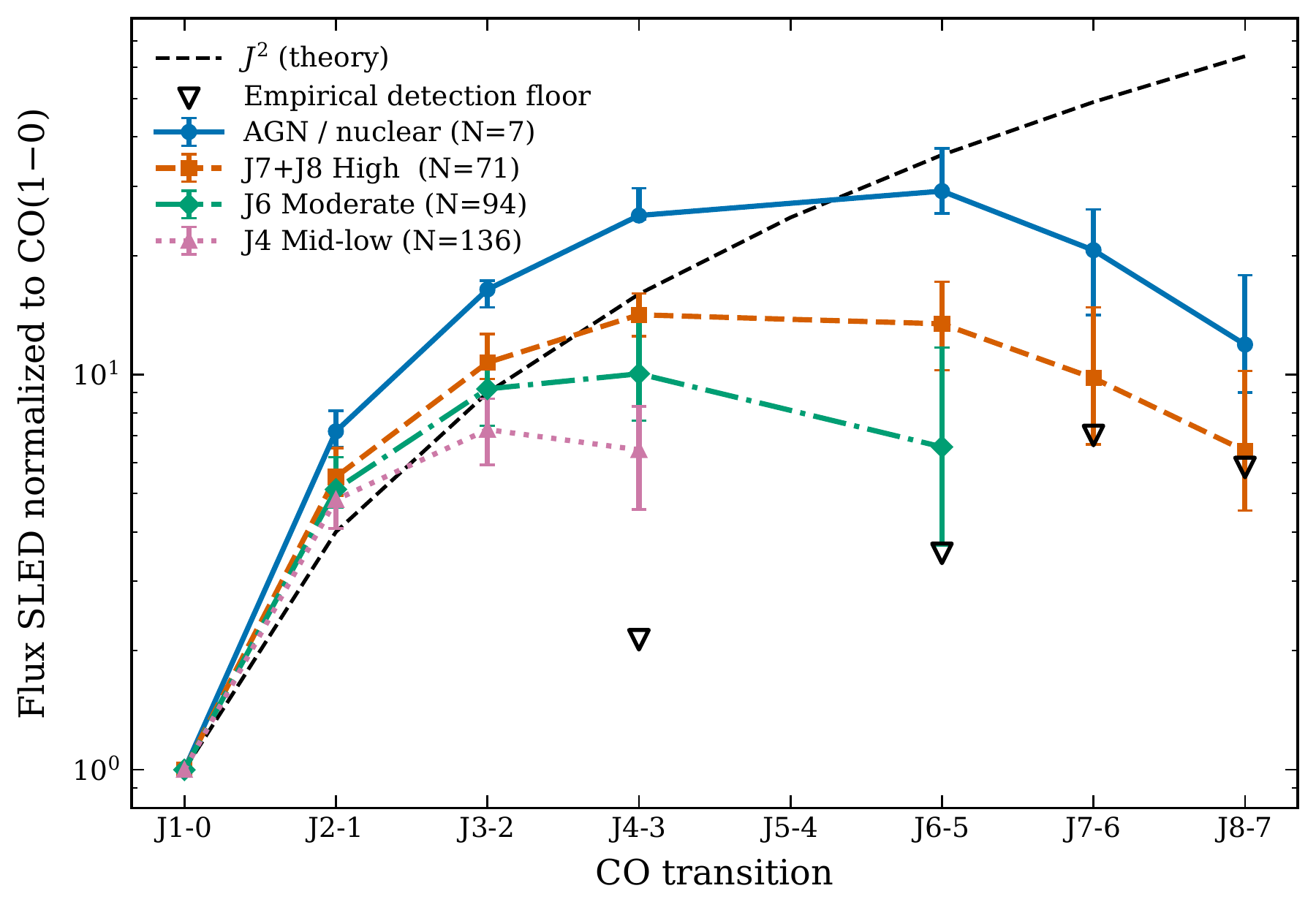}
\hfill
\includegraphics[width=0.51\textwidth]{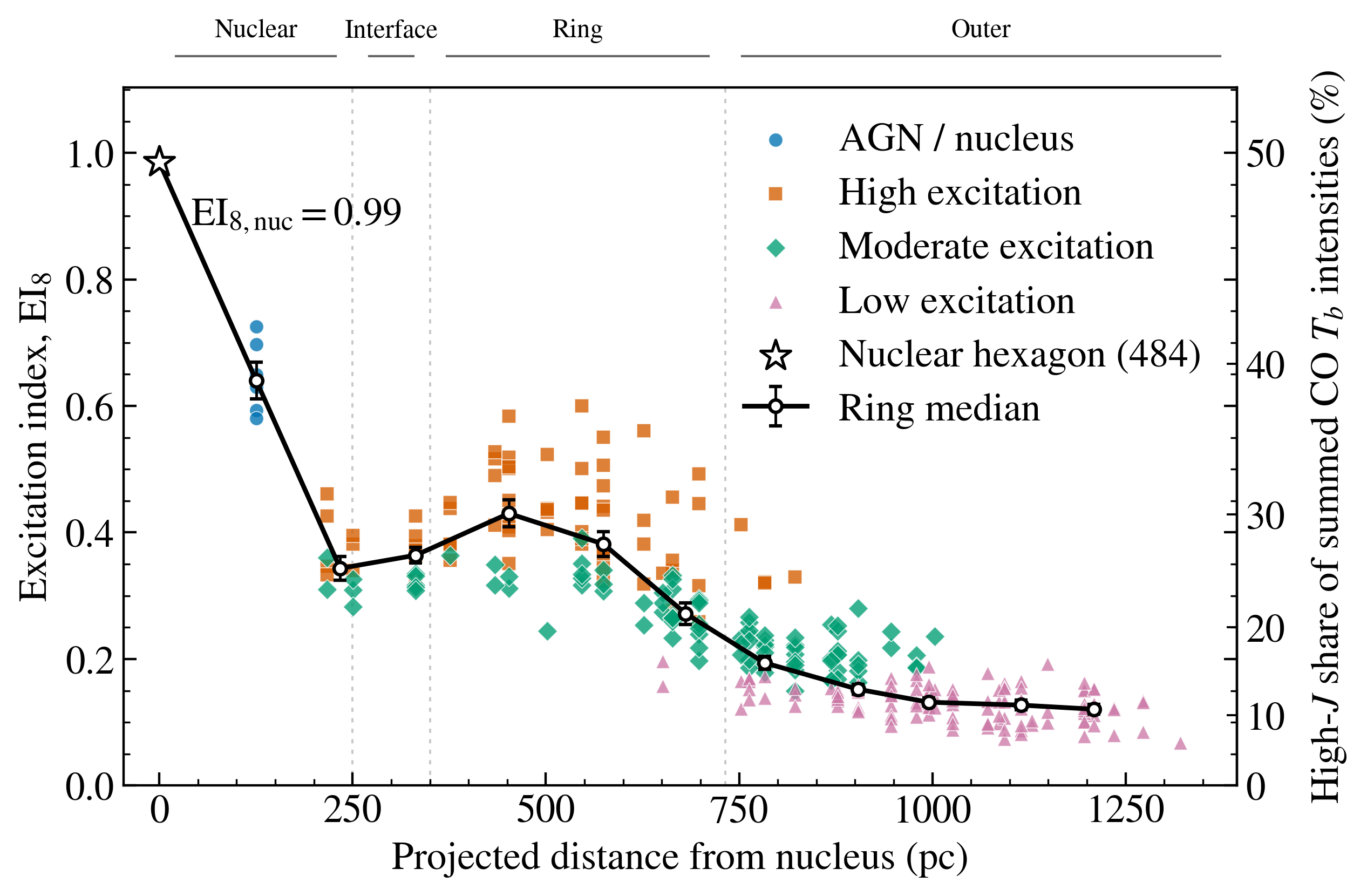}

\caption{
CO excitation structure in NGC~7469.
\textit{Top left:} Individual CO SLEDs normalized to CO(1--0), colored by CO(3--2) integrated intensity. Black lines show integrated CO SLEDs normalized by CO(1--0), using either all CO(1--0) detections in the selected sample (solid line) or only the hexagons jointly detected in CO(1--0) and each transition (dash-dotted line; see text). The gray dashed curve shows the $J^2$ scaling expected for optically thick, thermalized emission in the Rayleigh--Jeans limit, assuming a common emitting area. The open diamond marks the Herschel CO(5--4) measurement, obtained over a different aperture and shown for comparison.
\textit{Top right:} Excitation classes overlaid on the CO(1--0) integrated-intensity map. Colors distinguish the nuclear sample and the excitation classes defined by the highest detected transition (see text). The star marks the AGN position, and the ellipse indicates the synthesized beam.
\textit{Bottom left:} Median normalized flux SLEDs for the four excitation classes, with sample sizes indicated in the legend. Open downward triangles denote empirical detection floors; the dashed black curve shows the $J^2$ reference.
\textit{Bottom right:} Excitation index, EI$_{8}$, as a function of projected distance from the nucleus. Colored symbols identify the excitation classes, while connected black open circles show medians in successive hexagonal rings. The star marks the nuclear hexagon. The right-hand axis expresses the corresponding high-$J$ fraction of the summed CO brightness-temperature intensities. The nuclear, interface, ring, and outer zones are indicated above the panel.
}

\label{fig:CO_sled}
\end{figure*}

The top-left panel of Figure~\ref{fig:CO_sled} presents the resolved SLEDs normalized to CO(1--0). We define the parent sample as the 424 hexagons detected in all three lower-$J$ transitions: CO(1--0), CO(2--1), and CO(3--2). Detections in higher-$J$ transitions consequently select subsets of this sample. Individual SLEDs span a broad range of shapes above CO(3--2), demonstrating that a summed sequence does not fully describe the spatial variation in excitation.

The black curves compare two ways of constructing a summed CO SLED from the resolved measurements. The solid curve shows the total detected flux in each transition within the parent sample, normalized by the total CO(1--0) flux of that sample. In this case, because the detected area varies between transitions, the numerator and denominator generally integrate over different spatial regions. This resembles combining unresolved integrated measurements whose detected emission does not necessarily occupy the same area. The dash-dotted curve instead sums each transition and CO(1--0) over the same jointly detected hexagons, so that both fluxes refer to identical areas for each ratio.

The curves coincide through CO(3--2), where all parent-sample hexagons are detected, but diverge at higher $J$ as the detected coverage decreases. At CO(6--5), the normalized flux increases from $7.46$ for the solid curve to $12.01$ for the dash-dotted curve; at CO(8--7), it increases from $1.72$ to $8.91$. Thus, normalizing the compact higher-$J$ detections by CO(1--0) emission distributed over a larger area produces a more steeply declining summed SLED. The resolved measurements expose this spatial averaging: regions with substantial higher-$J$ emission can have a markedly different SLED from the sequence obtained by summing all detections in each transition. The Herschel CO(5--4) point is shown for comparison to guide the eye, but its different aperture prevents it from defining the missing transition in an individual hexagon or excitation class.

To examine the spatial distribution of different observed CO SLED shapes, we divide the resolved measurements into four empirical classes. Within the parent sample, we select the 308 hexagons also detected in CO(4--3). The central hexagon and its six immediate neighbors define the nuclear sample. The remaining positions are grouped according to their highest detected transition: the high-excitation class includes detections in CO(7--6) or CO(8--7); the moderate-excitation class includes CO(6--5) detections without either higher transition; and the lower-excitation class contains positions detected through CO(4--3) but without detections in the higher-$J$ lines. These samples contain 7, 71, 94, and 136 hexagons, respectively.

Their spatial distribution is shown in the top-right panel of Figure~\ref{fig:CO_sled}, and their median normalized SLEDs are compared in the bottom-left panel. Outside the nuclear sample, the high-excitation positions concentrate in bright regions of the circumnuclear ring, while the moderate- and lower-excitation positions extend farther across the surrounding molecular emission. The classification therefore provides a direct comparison between detected line coverage, SLED shape, and location.

The median normalized flux SLEDs in the bottom-left panel of Figure~\ref{fig:CO_sled} show progressively stronger separation above CO(3--2). The nuclear and high-excitation sequences remain approximately level between CO(4--3) and CO(6--5), before declining through CO(8--7). The nuclear median remains higher throughout this interval, exceeding the high-excitation median by approximately a factor of two at CO(6--5) and above. The moderate-excitation sequence instead decreases between CO(4--3) and CO(6--5), while the lower-excitation median already declines from CO(3--2) to CO(4--3). The sampled SLEDs therefore differ both in their normalization relative to CO(1--0) and in where their observed decline begins. The absence of CO(5--4) prevents a precise determination of the turnover between CO(4--3) and CO(6--5).

The plotted bars show the 16th--84th percentile spread among individual normalized SLEDs. Each median uses the detections available within its class, so the contributing positions can differ between transitions. The open downward triangles indicate empirical detection floors and do not constrain the undetected emission in a particular class.

To summarize the relative contributions of the observed lower- and higher-$J$ transitions in brightness-temperature units, we define

\begin{equation}
    \mathrm{EI}_{8}=
    \frac{\displaystyle\sum_{J\in\{4,6,7,8\}} W_J}
         {W_1+W_2+W_3},
    \label{eq:excitation_index}
\end{equation}

\noindent
where the subscript identifies the highest transition included, CO(8--7). The denominator combines CO(1--0), CO(2--1), and CO(3--2), the lower-$J$ lines commonly used in spatially resolved studies of molecular gas in nearby galaxies. The numerator sums the detected higher-$J$ transitions observed here, quantifying their combined brightness-temperature intensity relative to that traced by the more commonly observed lower-$J$ lines. $\mathrm{EI}_{8}$ summarizes the relative prominence of higher-$J$ emission in a single spatial diagnostic, allowing us to locate excitation enhancements and compare their radial behavior without reducing the analysis to one transition pair. It complements the individual line ratios and resolved SLEDs, which retain the transition-by-transition information compressed by the index.

All three denominator transitions must be detected, whereas only detected lines contribute to the numerator. CO(5--4) is not observed and is excluded from the definition. The associated fraction, $f_{\rm high}=\mathrm{EI}_{8}/(1+\mathrm{EI}_{8})$, expresses the contribution of the detected higher-$J$ lines to the summed included brightness-temperature intensities; it is not an energy-luminosity or cooling fraction. Undetected higher-$J$ emission would increase the index at fixed denominator. The measured values therefore provide an incomplete accounting of that emission, although differential spatial filtering prevents treating them as strict lower limits on the intrinsic ratios.

The bottom-right panel of Figure~\ref{fig:CO_sled} shows a central maximum and a secondary enhancement within the circumnuclear ring. The central nuclear hexagon has $\mathrm{EI}_{8}=0.985$, corresponding to $f_{\rm high}=0.496$, while its six immediate neighbors have a median of approximately $0.64$. The radial median decreases to $0.34$ near $230$~pc, rises to $0.43$ near $450$~pc, and declines beyond the ring to $0.12$--$0.15$ at approximately $0.9$--$1.2$~kpc. Connected points represent medians in successive hexagonal rings, plotted at the median projected radius of their contributing positions. The labeled zones are descriptive radial divisions.

Individual measurements show substantial variation at similar projected radii, particularly across the ring. The separation between excitation classes partly reflects their definition through higher-$J$ detections, which also contribute to the numerator. The radial profile consequently describes both the observed SLED variation and the changing line coverage.
\subsection{Nuclear and extra-nuclear contributions to the CO emission}
\label{sec:spatial_contributions}

\begin{figure}
\centering
\includegraphics[width=\columnwidth]{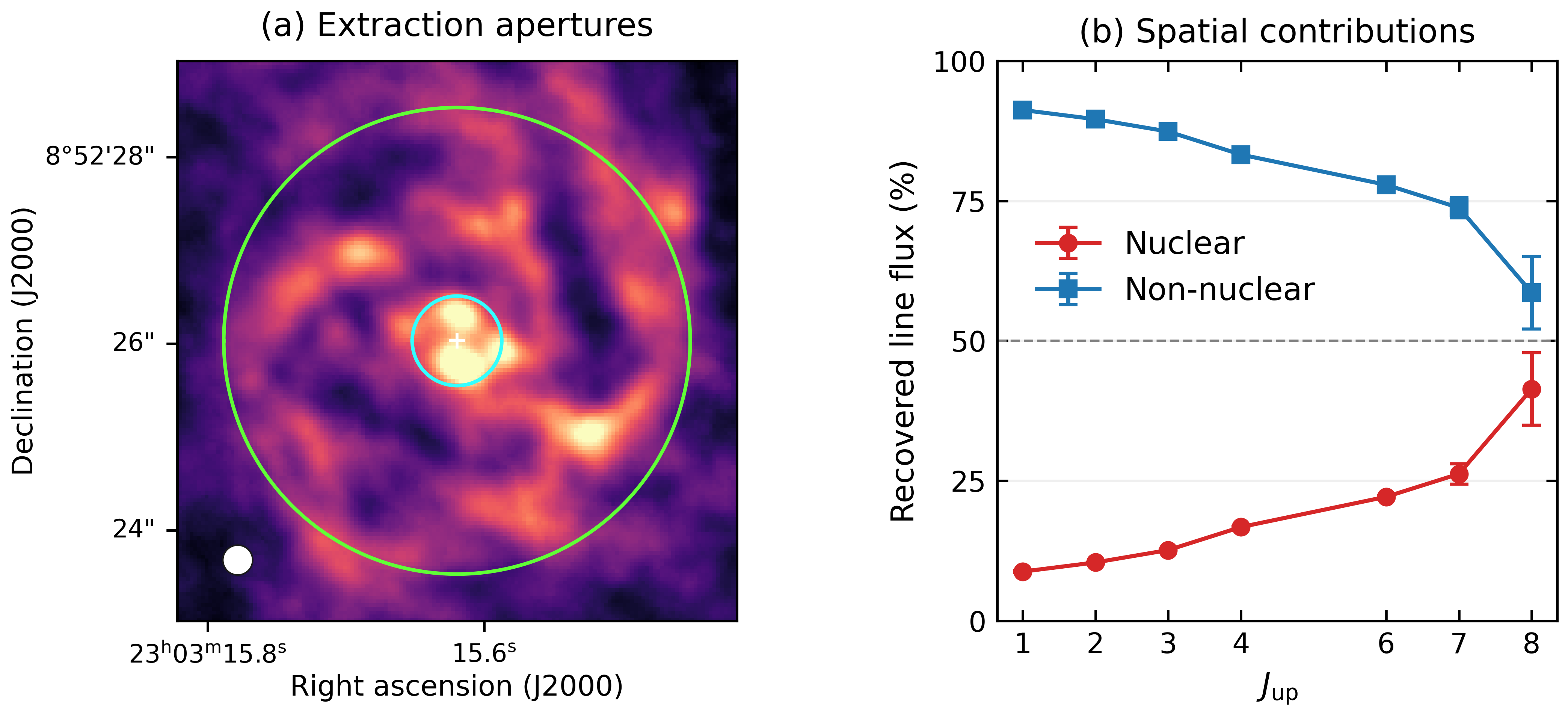}
\caption{Nuclear and extra-nuclear contributions to the recovered CO emission in NGC~7469. (a) Nuclear aperture ($r<0.48\arcsec$, cyan) and total extraction boundary ($r<2.50\arcsec$, green), overlaid on the CO(1--0) integrated-intensity map. The extra-nuclear region lies between these boundaries. The cross marks the center; the white circle shows the $0.32\arcsec$ beam. (b) Recovered line-flux fractions, with conditional statistical uncertainties. The dashed line marks equal contributions. All transitions use the common velocity interval defined in the text.}
\label{fig:nucleus_outside_co}
\end{figure}

We analyze the observed emission with a nuclear circular aperture of radius $0.48\arcsec$ ($\sim160$~pc; diameter $0.96\arcsec$, as adopted for the non-LTE analysis) and an annulus spanning $0.48<r<2.50\arcsec$ ($\sim160$--$825$~pc), both centered on the nucleus (Figure~\ref{fig:nucleus_outside_co}a). The outer radius was chosen to encompass the circumnuclear ring and the surrounding emission visible in the lower-$J$ maps. As discussed in Section~\ref{sec:totalfluxes}, these measurements characterize the recovered central emission over the sampled angular scales, rather than the complete molecular emission of the galaxy.

For this fixed-aperture decomposition, we integrate the spectra over the common interval $[-180,+250]\,\mathrm{km\,s^{-1}}$, relative to the redshifted transition frequencies at $z=0.01627$. This interval is covered by all seven cubes, including CO(8--7). These measurements are separate from sums over individually detected hexagons: no position-dependent detection cut is applied, and no absorption correction is made.

The total flux is measured within the circular aperture of radius $2.50\arcsec$ and satisfies $F_{\rm total}=F_{\rm nuc}+F_{\rm extra}$. We define $f_{\rm nuc}=F_{\rm nuc}/F_{\rm total}$ and $f_{\rm extra}=F_{\rm extra}/F_{\rm total}=1-f_{\rm nuc}$. These measurements are therefore distinct from sums over individually detected hexagons.

The fraction uncertainties are estimated from paired block-bootstrap resampling of the line-free aperture spectra, preserving covariance between the nuclear and extra-nuclear measurements. A common multiplicative calibration factor within each transition cancels in the fractions; the quoted errors exclude aperture and imaging systematics.

The nuclear fraction increases from approximately $9\%$ in CO(1--0) to $17\%$ in CO(4--3), $22\%$ in CO(6--5), and $26\%$ in CO(7--6) (Figure~\ref{fig:nucleus_outside_co}b). The extranuclear region therefore supplies approximately $74$--$91\%$ of the recovered flux from CO(1--0) through CO(7--6). In CO(8--7), the nominal nuclear and extra-nuclear fractions are approximately $41\%$ and $59\%$, respectively, with a shared conditional statistical uncertainty of about $6.5$ percentage points. The noise and aperture sensitivity of this transition prevent a firm assignment of dominance to either region.

The increasing nuclear fraction indicates greater central concentration of the recovered emission toward higher $J$. Its quantitative value nevertheless depends on the adopted spatial boundary and velocity interval. Differences in sensitivity and spatial-frequency sampling can also affect the relative recovery of compact and extended emission (Section~\ref{sec:totalfluxes}). The fractions therefore describe the spatial origin of the recovered line flux within the stated apertures, without identifying the relative contributions of different heating mechanisms. We discuss these results in Section~\ref{discussion_sub1}.

\subsection{Representative Aperture Analysis}
\label{sec:apertures}

The resolved measurements identify the nucleus and several bright ring complexes as regions with substantial emission beyond CO(3--2). The nucleus and three bright knots in the circumnuclear ring are detected in all seven observed CO transitions, providing complete SLEDs across the observed SLED from CO(1--0) to CO(8--7). We therefore focus our molecular-gas modeling on these four regions, using circular apertures of diameter $0.96\arcsec$, as detailed in Section~\ref{grid}. R1 encompasses the nucleus, while R2--R4 sample the three ring knots. These apertures are larger than the beam-sized hexagons and characterize the emission integrated over each region, rather than the conditions within individual grid elements. We use the R1--R4 identifiers consistently throughout the aperture analysis.

\subsubsection{Dust Continuum Spectral Energy Distributions}
\label{sec:dust_mbb}

\begin{figure*}[tp]
    \centering
    \includegraphics[width=0.8\textwidth]{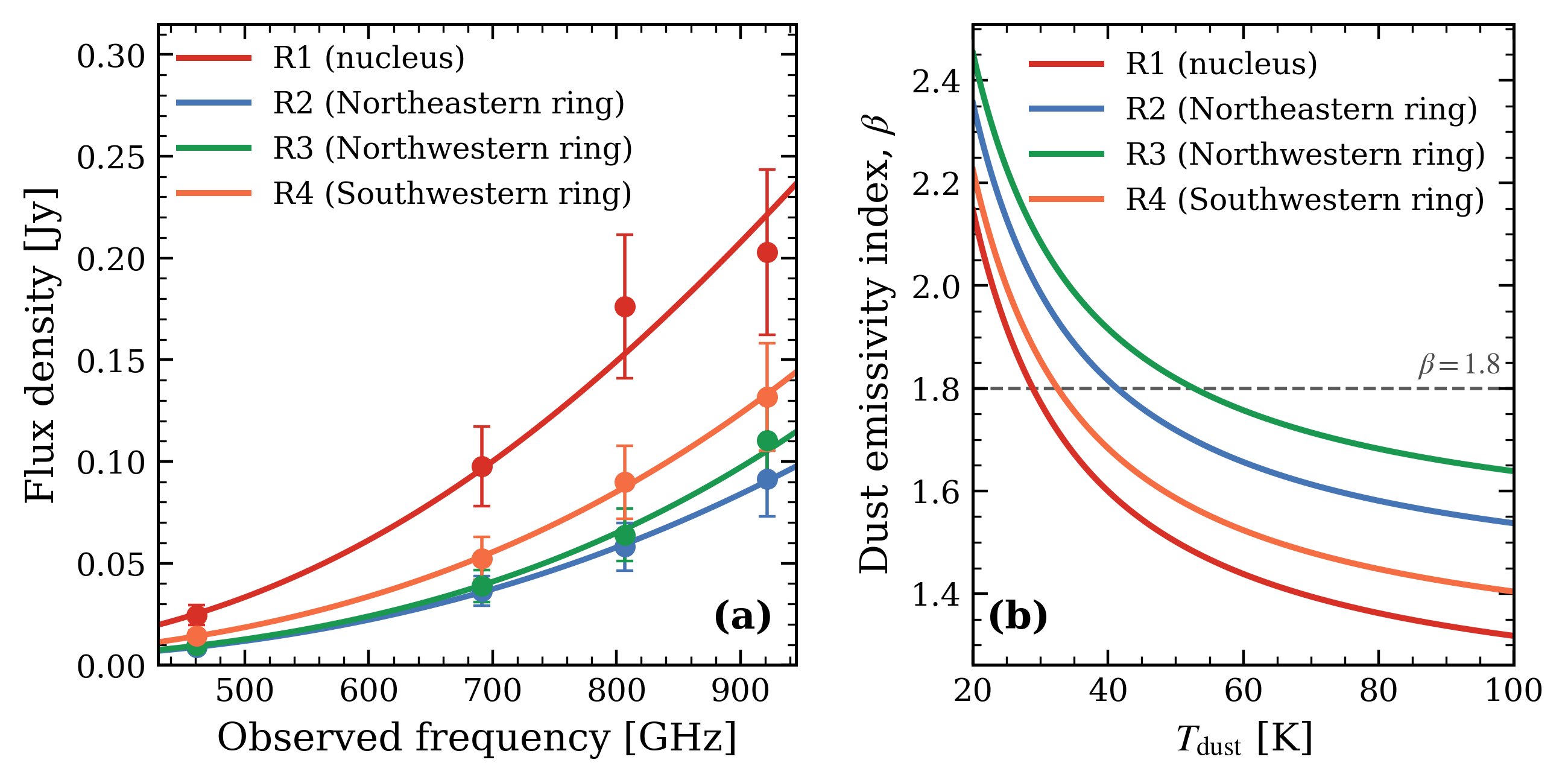}
    \caption{Aperture-integrated submillimeter continuum emission and modified-blackbody modeling in NGC 7469. (a) Measured flux densities for the nuclear aperture R1 and the three circumnuclear-ring apertures R2–R4, with vertical error bars indicating the adopted flux-density uncertainties. Solid curves show the modified-blackbody fits; colors identify the apertures as indicated in the legend. (b) Best-fitting dust emissivity index, ($\beta$), as a function of the assumed dust temperature, (T$_{\rm dust}$). The decreasing ($\beta$) with increasing temperature illustrates the temperature–emissivity degeneracy over the available frequency coverage. The horizontal dashed line marks ($\beta$=1.8) for reference.}
    \label{fig:knots}
\end{figure*}

Figure~\ref{fig:knots} presents the continuum spectral energy distributions of R1--R4 at approximately $461$, $691$, $807$, and $922$~GHz, with aperture flux densities listed in Table~\ref{tab:continuum_fluxes}. The nucleus is the brightest aperture at all four frequencies, followed by the southwestern ring aperture R4. The northeastern and northwestern apertures, R2 and R3, have comparable flux densities within their uncertainties. All four SEDs rise across the observed frequency range. The nuclear flux density increases from approximately $25$ to $203$~mJy between $461$ and $922$~GHz, while the ring apertures increase from $9$--$14$ to $91$--$132$~mJy. Between $807$ and $922$~GHz, the nuclear flux density rises by only approximately $15\%$, compared with $47$--$73\%$ in the ring apertures.

The modified-blackbody models reproduce the four-frequency SEDs, with minimum $\chi^2$ values of $0.67$, $0.02$, $0.11$, and $0.03$ for R1--R4, respectively. However, the fits admit a broad temperature--emissivity degeneracy, illustrated in the right-hand panel of Figure~\ref{fig:knots}. As the assumed dust temperature increases from $20$ to $100$~K, the fitted emissivity index decreases from approximately $2.1$--$2.5$ to $1.3$--$1.6$. The profile $\chi^2$ varies by less than $0.35$ across this entire interval in every aperture; consequently, none of the four SEDs provides a bounded temperature interval within the tested range at $\Delta\chi^2=1$.

For illustration, fixing $\beta=1.8$ yields conditional best-fit temperatures of approximately $28$, $41$, $55$, and $33$~K for R1--R4, respectively, consistent with the far-infrared dust temperatures inferred for nearby LIRGs \citep[e.g.,][]{Papadopoulos2000, Herrero-Illana_2019}. These values depend on the assumed emissivity index and do not establish a robust temperature ordering among the apertures. The continuum measurements therefore constrain the submillimeter spectral shapes and relative brightnesses more directly than the individual dust temperatures. We do not use these conditional temperatures as priors on the molecular-gas kinetic temperature.

\subsubsection{Aperture CO SLEDs}
\label{sec:aperture_sleds}

\begin{figure*}[tp]
    \centering
    \includegraphics[width=1\textwidth]{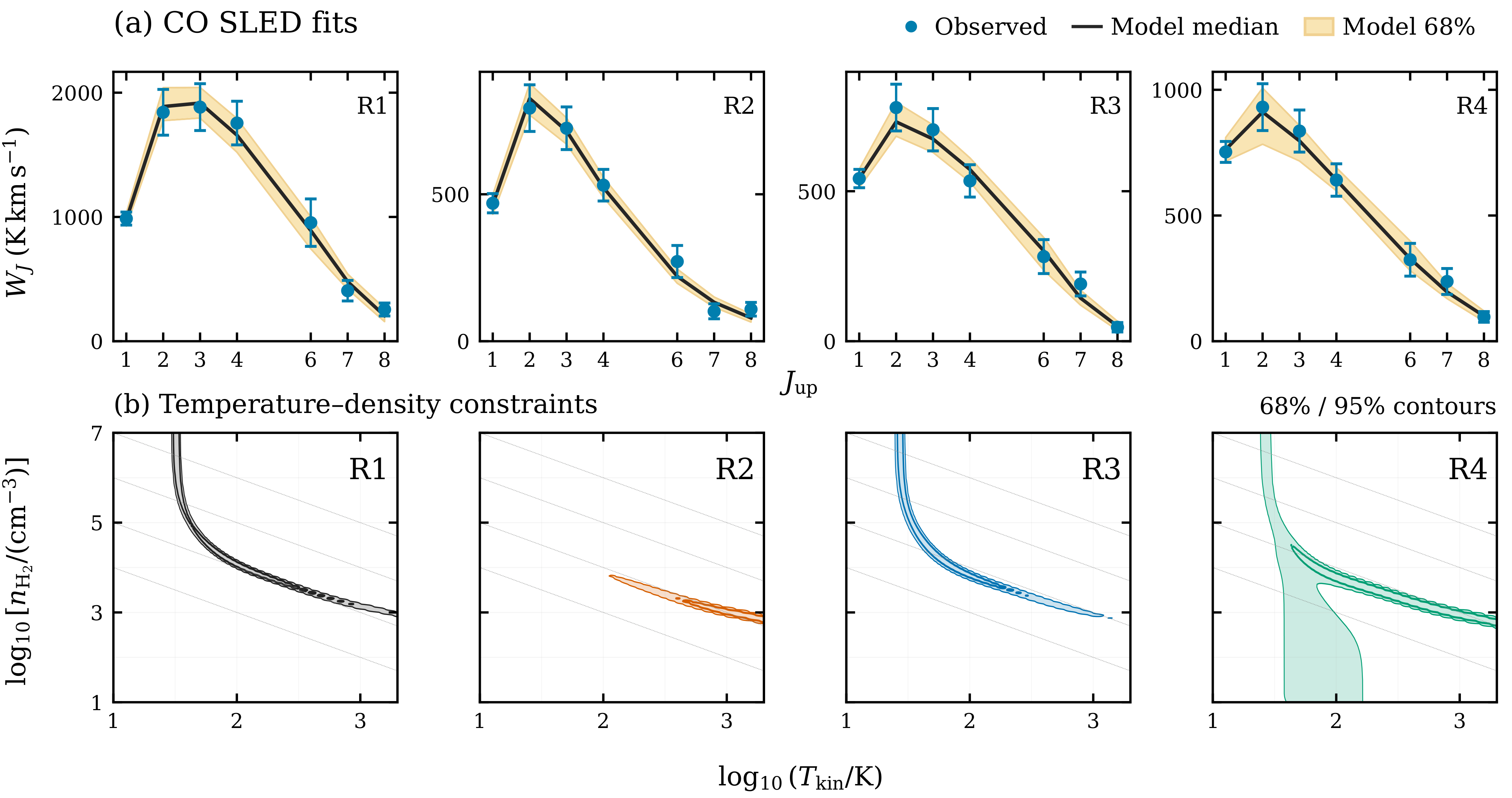}
    \caption{Fiducial one-component CO models for R1--R4. (a) Observed integrated brightness temperatures (points with adopted $1\sigma$ errors), pointwise posterior-median predictions (black curves), and 68\% model intervals (shading), including CO(8--7). (b) Joint temperature--density constraints, with 68\% and 95\% probability contours. Each column corresponds to the same aperture in both rows.}
    \label{fig:knots_sled}
    \label{fig:radex_joint_Tn}
\end{figure*}

Figure~\ref{fig:knots_sled}a compares the seven observed CO transitions with the fiducial one-component RADEX predictions for R1--R4. In contrast to the normalized flux SLEDs in Figure~\ref{fig:CO_sled}, these aperture measurements are expressed as integrated brightness temperatures. The nuclear aperture is brighter than the ring apertures throughout the observed CO SLED. Its brightness-temperature SLED rises from CO(1--0) to CO(2--1) and CO(3--2), remains elevated at CO(4--3), and declines toward CO(8--7). The ring SLEDs also turn over at low $J$ in these units, but differ in their relative upper-line strengths.

An effective one-component model provides a satisfactory description of each aperture under the adopted likelihood. The minimum grid $\chi^2$ values are $2.23$, $4.02$, $2.67$, and $1.11$ for R1--R4, respectively. For seven lines and four model parameters, these correspond to nominal reduced values of $0.74$, $1.34$, $0.89$, and $0.37$. The pointwise posterior-median line predictions differ from the observations by less than $1.3$ times the adopted uncertainty in every aperture. The ability to reproduce the observed intensities nevertheless does not imply that the model parameters are uniquely determined.

Table~\ref{tab:aperture_gas_properties} summarizes the marginalized parameters. The nuclear aperture R1 has a temperature median of $61$~K and a 16th--84th percentile interval of $33$--$338$~K, together with a much broader density distribution. The joint posterior admits cooler, denser solutions as well as warmer, more dilute alternatives. A comparable temperature--density degeneracy is present in R3, whose temperature interval is $27$--$157$~K. The substantial overlap between the R1 and R3 posteriors prevents a strong ranking of their temperatures or densities from their marginal medians alone. In particular, the strongly elevated nuclear CO ratios do not require the nucleus to have the highest kinetic temperature among the four fitted apertures.

\begin{table*}[tp]
\centering
\caption{Physical parameters of the fiducial one-component RADEX models. Values are marginalized posterior medians, with the 16th and 84th percentiles in brackets. Logarithms refer to numerical values in the units indicated. The pressure measure is calculated from joint posterior samples. These marginal summaries do not necessarily together describe a single representative model.}
\label{tab:aperture_gas_properties}
\small
\setlength{\tabcolsep}{5pt}
\begin{tabular}{lccccc}
\hline
Region & $T_{\rm kin}$ & $\log_{10}n_{\rm H_2}$ & $\log_{10}(N_{\rm CO}/\Delta v)$ & $\log_{10}(P_{\rm H_2}/k_{\rm B})$ & $\phi$ \\
 & (K) & ($\mathrm{cm^{-3}}$) & [$\mathrm{cm^{-2}\,(km\,s^{-1})^{-1}}$] & ($\mathrm{K\,cm^{-3}}$) & \\
\hline
R1 & $61\,[33,338]$ & $4.44\,[3.50,6.38]$ & $16.56\,[16.38,16.81]$ & $6.26\,[6.05,7.90]$ & $0.31\,[0.19,0.44]$ \\
R2 & $929\,[389,1567]$ & $3.06\,[2.88,3.31]$ & $16.19\,[16.00,16.50]$ & $6.01\,[5.91,6.10]$ & $0.44\,[0.25,0.70]$ \\
R3 & $43\,[27,157]$ & $4.56\,[3.63,6.25]$ & $16.81\,[16.63,17.00]$ & $6.18\,[5.84,7.69]$ & $0.37\,[0.25,0.52]$ \\
R4 & $187\,[53,808]$ & $3.31\,[2.81,3.94]$ & $17.00\,[16.56,18.06]$ & $5.84\,[5.36,5.99]$ & $0.25\,[0.19,0.43]$ \\
\hline
\end{tabular}
\end{table*}

R2 preferentially occupies a warmer, lower-density branch. Its temperature median is approximately $930$~K, with a 16th--84th percentile interval of $390$--$1570$~K, and its density is concentrated near $10^3~\mathrm{cm^{-3}}$. These values are conditional on the model and adopted priors: approximately $16\%$ of its temperature posterior lies above $1600$~K, approaching the upper prior boundary of $2000$~K. The data therefore favor a warm effective solution for R2 relative to R3, but do not provide a precise, prior-independent measurement of a temperature near $930$~K.

The joint molecular thermal-pressure measure is better constrained than temperature alone in R2, with
\begin{equation}
    \frac{P_{\rm H_2}}{k_{\rm B}}
    =n_{\rm H_2}T_{\rm kin}
    \simeq 1.03^{+0.24}_{-0.21}\times10^6~\mathrm{K\,cm^{-3}}.
\end{equation}
This improvement is not universal. In R1 and R3, high-density solutions remain compatible with the SLEDs over a broad range of density, producing extended upper tails in $n_{\rm H_2}T_{\rm kin}$. Their pressure distributions therefore remain broad and sensitive to the allowed density range.

R4 permits intermediate temperature solutions but retains a broad credible interval of approximately $53$--$808$~K. Its CO column density per unit linewidth is less well constrained than in the other apertures, with a 16th--84th percentile interval spanning approximately $1.5$~dex. This uncertainty reflects the limited separation of column density, excitation, and normalization within the adopted model; it does not by itself identify a particular optical-depth regime. The filling factors also remain effective brightness-scaling parameters and are not independently converted into emitting sizes.

The broad physical-parameter distributions persist after numerical refinement and direct RADEX validation, as described in Appendix~\ref{appendix:radex}. They therefore reflect the information available under the adopted physical model and priors, rather than being resolved by finer numerical sampling alone. The aperture comparison is most informative in the joint temperature--density distributions: the observed excitation differences cannot be interpreted as a simple sequence in kinetic temperature.

We also tested an additive two-component model using the same refined intrinsic spectra, observational uncertainties, and physical-parameter bounds. We compare the models using the Bayesian evidence, $Z$, defined as the likelihood integrated over the adopted parameter priors. We define $\Delta\ln Z\equiv\ln Z_{\rm two}-\ln Z_{\rm one}$, such that positive values favor the two-component model. Independent repeated fits yield $\Delta\ln Z$ between approximately $-0.04$ and $0.82$ across R1--R4, corresponding to Bayes factors $Z_{\rm two}/Z_{\rm one}=\exp(\Delta\ln Z)$ of approximately $0.96$--$2.26$. These values indicate no strong preference for the additional component. The added component improves the sampled fit quality, but the evidence provides only a weak preference, if any, for the more complex model. The separate component properties remain substantially degenerate, and the small evidence differences do not support ranking the apertures by their need for a second component. We therefore retain the one-component model as the fiducial description of the effective excitation conditions. This choice does not establish that the unresolved molecular medium is physically homogeneous or exclude multiple gas phases. In Section~\ref{discussion_sub3}, we explore whether an additional hotter contribution can account for the published mid-infrared H$_2$ emission while preserving the observed CO SLEDs; the illustrative two-component solutions and their diagnostics are presented in Appendix~\ref{app:h2_comparison}.

The joint temperature--density constraints are shown in Figure~\ref{fig:knots_sled}b; the numerical validation and model-comparison diagnostics are documented in Appendix~\ref{appendix:radex}.

\section{Discussion}
\label{discussion}
\subsection{Resolved excitation and the spatial origin of the integrated CO SLED}
\label{discussion_sub1}

Within the inner $\sim2$~kpc of NGC~7469, an active nucleus and a circumnuclear star-forming ring coexist within a rich molecular environment.  Our observations show that even though the strongest enhancement occurs toward the nucleus, the circumnuclear structures are important sources of excited CO emission. The ring is not simply a lower-excitation background surrounding a distinct nuclear source: bright ring complexes have upper CO SLED slopes comparable to the nucleus, and the surrounding molecular structures collectively supply most of the recovered emission through CO(7--6) (Figures~\ref{fig:CO_sled} and~\ref{fig:nucleus_outside_co}). The distinction matters because a spatially integrated SLED reflects both the local excitation and the brightness and spatial extent of the emitting gas. A bright, highly excited nucleus does not necessarily determine the shape of the integrated SLED if less excited emission occupies a substantially larger area.

The ratio maps reveal an excitation structure that is neither uniform within the ring nor determined by projected distance from the nucleus alone (Figure~\ref{fig:ratios}). The central enhancement is accompanied by localized variations, with some circumnuclear positions attaining consecutive-transition ratios comparable to or higher than the nuclear values. Similar relative strengths between successive transitions can therefore coexist with different strengths relative to the ground transition. The apparent nuclear--ring contrast consequently depends on which transitions are compared.

The comparison between $R_{21}$, $R_{32}$, and $R_{43}$ helps locate where the nuclear enhancement develops. In the central hexagon, CO(2--1) is approximately twice as bright as CO(1--0) in integrated brightness-temperature units, while $R_{32}$ and $R_{43}$ remain close to unity. The elevated $R_{31}$ and $R_{41}$ consequently preserve the contrast already established between CO(2--1) and CO(1--0), rather than indicate a similar increase at every successive transition. Nevertheless, the central enhancements in $R_{32}$ and $R_{43}$ relative to the surrounding emission show that the nuclear distinction is not confined to the ground transition. Maintaining these near-unity consecutive ratios is itself a distinction from much of the outer disk, where their lower values indicate a decline in integrated brightness temperature through CO(3--2) and CO(4--3).

At the highest observed transitions, the patchier $R_{76}$ and $R_{87}$ distributions show that the nucleus is less distinctive in the relative strengths of these lines than in their ratios to CO(1--0). This behavior gives a physical basis for distinguishing CO(1--0)-normalized ratios from consecutive-transition ratios. CO(1--0) can remain bright in relatively cool molecular material that contributes little to the upper transitions. Varying the contribution of such material changes the upper-to-CO(1--0) ratios without necessarily producing a corresponding change between successive higher transitions. The observed contrasts are therefore consistent with different contributions from molecular material capable of sustaining upper-transition emission. They do not, however, demonstrate separate gas components. Likewise, ratios above unity relative to CO(1--0) should not be identified directly with superthermal excitation. Unity is the reference for optically thick emission at a common excitation temperature in the Rayleigh--Jeans limit, with equal emitting areas and velocity profiles. Departures from these assumptions, including transition-dependent optical depths, can produce different ratios without requiring superthermal level populations \citep{RADEX}. The maps establish spatially varying CO SLED shapes; they do not uniquely assign a temperature, density, or heating mechanism to those variations. CO excitation modeling of the nucleus and three selected ring complexes is presented in Section~\ref{sec:aperture_sleds}, with its physical interpretation discussed in Section~\ref{discussion_sub2}.

The $\mathrm{EI}_{8}$ distribution places these line-ratio variations in a common spatial context (Figure~\ref{fig:CO_sled}). By comparing the summed detected upper-transition intensities with the lower CO(1--0), CO(2--1), and CO(3--2), it measures their relative prominence without relying on the ground transition alone. The nuclear maximum therefore shows that the central enhancement persists when the reference includes all three lower transitions. It does not imply that every successive ratio is unusually high, but that the included upper emission is collectively stronger relative to the lower portion of the SLED.

The radial distribution reveals both a broad organization and substantial departures from it. The decline outside the nucleus is interrupted by a secondary enhancement within the ring, where some hexagons reach values comparable to those immediately surrounding the central nuclear hexagon. Other ring positions at similar projected radii have appreciably lower values. The scatter is therefore as informative as the radial median: elevated relative upper-transition emission occurs locally within the ring, rather than forming a uniform annulus or following projected distance from the AGN alone.

Comparable $\mathrm{EI}_{8}$ values nevertheless do not require identical SLED shapes, because different combinations of line strengths can produce the same summed ratio. The index identifies where upper-transition emission is prominent, while the individual ratios distinguish how that emission is distributed among transitions. The set of upper transitions entering the index also varies between positions: an undetected line contributes no measured intensity to the numerator. Differences in sensitivity and line coverage therefore contribute to the observed spatial pattern. The detection classes are consequently not independent evidence for distinct physical conditions. Read together with the ratio maps, $\mathrm{EI}_{8}$ supports a central excitation enhancement with localized circumnuclear variations, without assigning a unique temperature or heating mechanism to either.


\subsection{Molecular gas excitation in the nucleus and star-forming ring}
\label{discussion_sub2}

The resolved observations establish two complementary features: the nuclear CO SLED is strongly enhanced relative to CO(1--0), while bright ring complexes retain substantial upper-transition emission (Sections~\ref{sec:ratios} and~\ref{sec:sleds}). To interpret these results, we adopt the fiducial one-component non-LTE models described in Section~\ref{sec:radex_method} for the nucleus and three bright ring knots with detections in all seven transitions.

The one-component models provide a physically motivated explanation of the excitation diversity described above: the stronger nuclear CO enhancement does not require a higher kinetic temperature than in the ring. The joint temperature--density distributions allow different balances between collisional excitation and radiative escape to account for their SLEDs (Figure~\ref{fig:knots_sled}a-b). The comparison is therefore most informative in terms of these alternative physical conditions, rather than a temperature sequence inferred from the relative strength of the upper CO emission.

In the nucleus, the dense branch offers an explanation for sustaining substantial upper-transition emission at moderate kinetic temperatures. Frequent collisions populate the rotational levels, while photon trapping reduces the effective radiative decay rate \citep{RADEX}. The nuclear concentration of HCN(4--3), HCO$^+$(4--3), and CS(7--6) reported by \citet{Izumi2015} makes this interpretation physically plausible. Those tracers do not select a unique density for our CO-emitting component, but they provide independent support for dense material in the environment where our models permit it. The warmer, lower-density alternatives remain viable, so the nuclear SLED supports efficient excitation without identifying a unique combination of temperature and density.

This interpretation is compatible with a thermally and chemically structured nuclear medium. The enhanced [C\,\textsc{I}] emission and neutral-carbon-to-CO abundance inferred by \citet{Izumi_2020} favor an X-ray-affected environment, with inferred temperatures overlapping the warmer part of our nuclear distribution. Our CO-only constraints neither reproduce that chemical diagnostic nor exclude its interpretation. Rather, they allow dense molecular material to contribute strongly to the CO SLED while gas more affected by the AGN contributes differently to other tracers. Indeed, the nuclear H$_2$ rotational emission detected by \citet{Armus_2023} independently establishes a warm molecular contribution. Together, these observations motivate a range of molecular conditions within the nuclear aperture. The one-component CO description remains useful because it accounts for the measured CO emission; its temperature is not a census of all the gas exposed to the AGN. The independent warm-gas signatures therefore raise a question about the extent of that description, rather than invalidate its fit to CO. The possibility that CO and H$_2$ emphasize different molecular conditions is explored for the ring apertures through the illustrative two-component comparison in Section~\ref{discussion_sub3}.

The ring poses the complementary question: why can a complex outside the nucleus favor a warmer solution despite its less elevated CO SLED relative to CO(1--0)? R2 illustrates this case. At its preferred lower density, fewer collisions occur per molecule, and higher temperature helps sustain the observed excitation. The contrast with the cooler, denser solutions allowed in R3 therefore suggests that the ring SLED differences can reflect different balances of temperature and density, rather than simply different degrees of heating along one common sequence. R4's broader distribution does not establish a further step in that ordering. For R2, the concentration of the joint solutions around $n_{\rm H_2}T_{\rm kin}\sim10^6~\mathrm{K~cm^{-3}}$ is useful because it characterizes the favored excitation conditions more consistently than temperature alone. It does not establish pressure equilibrium with the other complexes or include their turbulent and magnetic support.

The local infrared environment makes warm molecular material in these ring apertures plausible. R2--R4 overlap LR7, LR9, and LR1 of \citet{Lai_2022} and contain the MIRI sources M2, M3, and M1 of \citet{Bohn_2023}, respectively (Table~\ref{tab:region_co_constraints}). Their association with embedded star formation, together with the ring's PAH and H$_2$/PAH diagnostics, makes stellar irradiation a natural explanation for warm molecular material in these complexes, consistent with the warm solution favored for R2. Irradiated cloud surfaces and shielded interiors can contribute differently to the observed CO transitions \citep{Joblin_2018}; changing their relative contributions within an aperture can consequently change its effective one-component solution. This offers a physical interpretation of the warm R2 solution without assigning its fitted temperature to the entire molecular reservoir.

Mechanical heating may supplement this picture locally, but the spatial evidence matters. The shock-sensitive line ratios and enhanced dispersions identified by \citet{Vivian_2022} concern gas between the nucleus and ring, approximately $180$~pc west and north of the AGN. They demonstrate that mechanical energy reaches some circumnuclear molecular material; they do not establish shock heating in all three modeled ring apertures. Thus, the warm solutions are consistent with an environment containing both stellar heating and localized dissipation, while their attribution to a particular energy source must follow the corresponding environmental diagnostics.

The continuum provides no independent temperature ordering with which to interpret the CO differences. Whereas the CO SLEDs distinguish the excitation across the apertures, the continuum primarily distinguishes their brightnesses. The temperature--emissivity degeneracy allows similar dust spectral shapes over a broad range of temperatures, so the continuum fits cannot determine whether the different effective CO solutions are accompanied by corresponding differences in dust temperature.

\begin{table}[t]
\centering
\caption{Associations with published JWST regions.}
\label{tab:region_co_constraints}
\small
\setlength{\tabcolsep}{4pt}
\begin{tabular}{lccc}
\hline\hline
Aperture & Lai et al. ID & Bohn et al. ID & Offset ($\arcsec$) \\
\hline
R1 & Nucleus & --- & --- \\
R2 & LR7 & M2 & 0.13 \\
R3 & LR9 & M3 & 0.21 \\
R4 & LR1 & M1 & 0.18 \\
\hline
\end{tabular}
\tablecomments{
LR identifies regions from \citet{Lai_2022};
M identifies MIR sources from \citet{Bohn_2023}.
Offsets separate the Bohn et al. sources from our aperture centers.
These are associations between overlapping complexes.
Our R1 is the nucleus; LR1 lies in the southwestern ring.
The CO apertures have $0.96\arcsec$ diameters;
the CO--H$_2$ comparison uses $0.8\arcsec$ apertures
at the Lai et al. centers.
}
\end{table}

\subsection{The molecular temperature distribution traced by CO and H$_2$}
\label{discussion_sub3}

The preceding interpretation describes the conditions capable of producing the CO emission within each aperture, but leaves open how much unresolved structure those effective conditions represent. The temperature--density alternatives allowed by a one-component fit are different explanations of the measured SLED; they do not establish that the corresponding gas components coexist. Our CO-only two-component models explore whether the seven observed transitions provide evidence for separating the emission into two contributions with different physical conditions.

Although the additional component improves the sampled fit quality, the Bayesian evidence provides no compelling preference for that description under the adopted priors (Table~\ref{tab:appendix_co_validation}). The extra flexibility therefore does not yield a well-established physical partition of the CO emission. This supports retaining one component as the fiducial description, while limiting the interpretation of its parameters to the emission sampled by these transitions. In particular, the CO-only comparison cannot establish whether molecular material contributing weakly through CO(8--7) is present within the same aperture.

Mid-infrared H$_2$ emission offers a way to explore that possibility because its rotational lines probe substantially higher upper-level energies than the CO transitions observed here \citep{TogiSmith_2016}. We therefore compare CO-based descriptions with the published H$_2$ measurements of \citet{Lai_2022} in the ring regions LR7, LR9, and LR1, using CO extractions in their nominal comparison apertures. This extends the physical question beyond whether two components improve the CO fit: it examines whether the conditions accounting for CO can also supply the observed higher-energy molecular emission.

We find that the tested homogeneous descriptions underpredict the higher-energy H$_2$ lines (Appendix~\ref{app:h2_comparison}; Figure~\ref{fig:h2_component_comparison}). Our illustrative two-component solutions offer a possible physical explanation: their hotter contribution has a smaller effective filling factor, $\phi$, than the cooler contribution dominating the CO emission. This is consistent with a picture in which localized hot molecular regions lie within a more extended, cooler CO-emitting medium. Their high temperature allows them to contribute strongly to the higher-energy H$_2$ lines without requiring the same conditions throughout the material producing most of the observed CO SLED.

The component parameters and fit diagnostics are given in Appendix~\ref{app:joint_examples}. Their agreement with the H$_2$ measurements depends on the adopted excitation assumptions and illustrative uncertainty allowance. Moreover, $\phi$ scales the modeled brightness rather than independently measuring an emitting size, so this spatial arrangement is a plausible interpretation rather than an inferred geometry. These exploratory examples demonstrate how the two tracers could emphasize different molecular structures; they do not establish a unique physical decomposition. Spatially matched CO and H$_2$ observations at higher resolution, together with additional CO transitions above CO(8--7), could help test this interpretation.

Two discrete components are one representation of that structure. Temperature and density gradients within irradiated clouds, or localized dissipation, can also produce a range of conditions within an aperture \citep{Joblin_2018,NeufeldYuan_2008}. Continuous temperature distributions have likewise been used to describe molecular hydrogen emission in galaxies \citep{TogiSmith_2016,Lai_2022}. Our comparison does not select a unique distribution; it shows why the effective conditions describing the observed CO SLED should not be assigned to all molecular material within the same region.

\section{Conclusions}
\label{sec}

We have examined molecular gas excitation across the central $\sim$2 kpc of NGC~7469, encompassing the nucleus and circumnuclear star-forming ring, using matched-resolution ALMA observations of seven CO transitions from CO(1--0) to CO(8--7) at a spatial resolution of $\sim$120 pc. Our principal conclusions are as follows.

\begin{enumerate}

\item The strongest CO excitation enhancement occurs toward the nucleus, but the ring contains substantial local variations that cannot be described by projected distance from the AGN alone. Some ring positions attain highest consecutive-transition ratios comparable to or greater than the nuclear values. The secondary enhancement and scatter in $\mathrm{EI}_{8}$ show that the relative prominence of upper-transition emission varies within the ring, with some positions resembling the immediate nuclear surroundings.

\item The nuclear enhancement depends on which transitions are compared. In the central hexagon, $R_{21}=1.99\pm0.23$, while $R_{32}$ and $R_{43}$ remain near unity. The elevated $R_{31}$ and $R_{41}$ therefore retain much of the contrast established between CO(2--1) and CO(1--0), rather than representing a comparable increase at every successive transition. Ratios above unity relative to CO(1--0) do not uniquely establish superthermal excitation; their interpretation also depends on optical depth, excitation temperature, and the emitting structures contributing to each transition.

\item Stronger nuclear excitation does not imply nuclear dominance of the recovered emission. Within the central $\sim1.65$~kpc-diameter field, the extra-nuclear annulus supplies approximately $74$--$91\%$ of the recovered flux from CO(1--0) through CO(7--6), despite an increasing nuclear contribution toward higher transitions. The spatial partition in CO(8--7) remains less certain. These measurements include the ring and the regions between the ring and nucleus, and describe spatial contributions rather than fractions attributable to AGN or stellar heating.

\item The nuclear and ring SLEDs permit different balances of molecular density and temperature. Dense gas
($n_{\mathrm{H}_2} \sim 10^{4.4}\,\mathrm{cm}^{-3}$) provides a plausible explanation for the nuclear excitation, while warmer, lower-density solutions also remain allowed. The northeastern ring complex R2 favors a warm, relatively low-density solution ($T_{\rm kin} = 930^{+640}_{-540}$ K, $n_{\mathrm{H}_2} \sim 10^3\,\mathrm{cm}^{-3}$), with $n_{\mathrm{H}_2}T_{\rm kin} \sim 10^6\,\mathrm{K\,cm}^{-3}$, whereas R3 admits cooler, denser conditions. The strongest enhancement relative to CO(1--0) therefore does not identify a unique kinetic-temperature ordering among the regions.

\item The contrast in CO excitation has no comparably clear counterpart in the submillimeter continuum spectral shapes. Across $461$--$922$ GHz, the continuum primarily distinguishes the aperture brightnesses, with the nucleus brightest throughout. Fixing $\beta=1.8$ gives conditional dust temperatures of approximately $28$--$55$ K, but the temperature--emissivity degeneracy prevents establishing a dust-temperature ordering corresponding to the CO differences.

\item In the ring comparison apertures, the tested homogeneous CO-based descriptions underpredict the higher-energy mid-infrared H$_2$ emission reported in the literature. Exploratory two-component models show that an additional hot molecular component ($T_{\rm kin} \sim 800$ K), with a smaller effective filling factor than the cooler CO-emitting component, can account for this emission while contributing less than 10\% of the modeled CO(8--7) flux. These solutions depend on the adopted excitation assumptions and uncertainty allowance and do not establish a unique physical decomposition. The comparison illustrates how an effective one-component CO description can leave additional hot molecular gas weakly represented.

\end{enumerate}

NGC~7469 demonstrates that the location of the strongest CO excitation differs from the spatial origin of most of the recovered CO emission. An unresolved SLED combines the brightness and extent of different environments, while individual molecular tracers can emphasize different conditions within those environments. Its shape therefore cannot uniquely specify a molecular-gas temperature or the contribution of AGN heating.

\section*{Acknowledgements}
This paper makes use of the following ALMA data: ADS/JAO.ALMA$\#$2017.1.00078.S, ADS/JAO.ALMA$\#$2023.1.00499.S. ALMA is a partnership of ESO (representing its member states), NSF (USA) and NINS (Japan), together with NRC (Canada), NSTC and ASIAA (Taiwan), and KASI (Republic of Korea), in cooperation with the Republic of Chile. The Joint ALMA Observatory is operated by ESO, AUI/NRAO and NAOJ.

Support for this work was provided by the NSF through the Grote Reber Fellowship Program administered by Associated Universities, Inc. National Radio Astronomy Observatory; ALMA Student Observing Support (SOS) program. A.S. would like to thank Dr. Annalisa Pillepich (Max Planck Institute for Astronomy) and Dr. Dylan Nelson (University of Heidelberg) for insightful discussions on the theory of galaxy evolution and excitation mechanisms. VU acknowledge funding support from NSF Astronomy and Astrophysics Research Grant $\#$2536603, NASA Astrophysics Decadal Survey Precursor Science grant $\#$80NSSC25K0169, and JWST-GO-01717.001-A. CML acknowledges that this research project was supported by the Hellenic Foundation for Research and Innovation (HFRI) under the ``2nd Call for HFRI Research Projects to support Faculty Members \& Researchers'' (Project Number: 03382). AMM acknowledges support from the NASA Astrophysics Data Analysis Program (ADAP) grant number 80NSSC23K0750, from NSF AAG grant $\#$2009416 and NSF CAREER grant $\#$2239807, and from the Research Corporation for Science Advancement (RCSA) through the Cottrell Scholars Award CS-CSA-2024-092. MB acknowledges support from the Juan de La Cierva scholarship with reference JDC2023-052684-I, funded by MICIU/AEI/10.13039/501100011033 and FSE+ and from the Agencia Estatal de Investigaci\'on of the Ministerio de Ciencia, Innovaci\'on y Universidades (MCIU/AEI) under the grant ``Tracking active galactic nuclei feedback from parsec to kiloparsec scales'', with reference PID2022$-$141105NB$-$I00 and the European Regional Development Fund (ERDF). ET gratefully acknowledges financial support from ANID CATA-BASAL - FB210003, and FONDECYT Regular - 1241005 and 1250821. NS acknowledges support provided by the NSF through the Grote Reber Fellowship Program administered by Associated Universities, Inc./National Radio Astronomy Observatory. During the preparation of this manuscript, the authors used ChatGPT (OpenAI) to assist with language editing and formatting. The authors reviewed and revised all AI-assisted content and take full responsibility for the final manuscript.

\clearpage
\appendix
\raggedbottom
\section{CO-model inputs and validation}
\label{appendix:radex}

We document the CO-model inputs and numerical validation, followed by supplementary nuclear and CO--H$_2$ comparisons. The fiducial CO fits use $0.96\arcsec$-diameter apertures; the H$_2$ comparison uses separate CO extractions in nominal $0.8\arcsec$ apertures.

\subsection{Molecular inputs and model parameters}
\label{app:radex_inputs}

CO intensities are calculated with RADEX \citep{RADEX} through SpectralRadex 1.1.5, using the packaged \texttt{co.dat}: 41 rotational levels and separate CO--para-H$_2$ and CO--ortho-H$_2$ rates from \citet{Yang_2010}, tabulated over 2--3000~K. We supply the total H$_2$ density and use the backend's default thermal treatment of the collision partners. Atomic hydrogen and electron densities are set to zero, and the only background radiation is a 2.73~K blackbody.

We adopt the uniform-sphere escape probability (\texttt{geometry=1}).\footnote{\url{https://spectralradex.readthedocs.io/en/latest/api/radex.html}} Table~\ref{tab:radex_grid} summarizes the fixed inputs and priors.

\begin{table}[!h]
\centering
\caption{CO-model parameters, priors, and fixed inputs.}
\label{tab:radex_grid}
\small\setlength{\tabcolsep}{5pt}
\begin{tabular}{ll}
\hline
Quantity & Adopted specification \\
\hline
$T_{\rm kin}$ & $10$--$2000$~K; uniform in $\log_{10}T_{\rm kin}$ \\
$n_{\rm H_2}$ & $10^1$--$10^7\,\mathrm{cm^{-3}}$; uniform in $\log_{10}n_{\rm H_2}$ \\
$Q\equiv N_{\rm CO}/\Delta v$ & $10^{15}$--$10^{21}\,\mathrm{cm^{-2}\,(km\,s^{-1})^{-1}}$; uniform in $\log_{10}Q$ \\
$\phi$ & Uniform on $[0,1]$ per component \\
$\Delta v$ (FWHM) & R1: 190; R2: 82; R3: 61; R4: 71~$\mathrm{km\,s^{-1}}$ \\
$N_{\rm CO}$ in backend & $Q\Delta v$, using the fixed regional FWHM \\
Escape probability & Uniform sphere, \texttt{geometry=1} \\
Background & $T_{\rm bg}=2.73$~K \\
Collisions & H$_2$, using the packaged ortho/para rate data \\
Intrinsic cache & 39 temperature nodes; 25 density nodes; 25 column-per-velocity nodes \\
Refinement coordinates & $\log_{10}T_{\rm kin}$, $\log_{10}n_{\rm H_2}$, $\log_{10}Q$ \\
\hline
\end{tabular}
\tablecomments{Temperature nodes are logarithmically spaced from 10 to 2000~K. The original density and column axes have 0.25-dex spacing. The linewidths are held fixed within each regional analysis; the same regional value applies to both components of a two-component model. The filling factors scale line brightness and are not independently converted into source sizes or constrained to sum to unity.}
\end{table}

The Gaussian intensity likelihood uses the statistic in Equation~\ref{eq:radex_chi2_method}, with independent measurement and calibration errors. The CO-only fits impose no abundance or dust-temperature constraint. Below, $m_i(\boldsymbol{\theta})$ denotes the intrinsic integrated intensity for $\boldsymbol{\theta}=(\log_{10}T_{\rm kin},\log_{10}n_{\rm H_2},\log_{10}Q)$, and $y_i$ and $\sigma_i$ are the measured intensity and its total uncertainty.

\subsection{Marginalization, numerical refinement, and model comparison}
\label{app:radex_inference}

For the one-component model, the uniform filling-factor prior permits analytic marginalization. Define
\begin{equation}
 A=\sum_i\frac{m_i^2}{\sigma_i^2},\qquad
 B=\sum_i\frac{m_i y_i}{\sigma_i^2},\qquad
 C=\sum_i\frac{y_i^2}{\sigma_i^2}.
\end{equation}
Then, omitting the common data-dependent normalization,
\begin{align}
 \overline{\mathcal{L}}(\boldsymbol{\theta})
 &=\int_0^1\mathcal{L}(\boldsymbol{\theta},\phi)\,d\phi \nonumber\\*
 &=\exp\left[-\frac{1}{2}\left(C-\frac{B^2}{A}\right)\right]
 \sqrt{\frac{2\pi}{A}}\,
 \left[\Phi\!\left(\sqrt{A}\left[1-\frac{B}{A}\right]\right)
 -\Phi\!\left(-\frac{B}{\sqrt{A}}\right)\right],
 \label{eq:app_marginal_phi}
\end{align}
where $\Phi$ is the standard-normal cumulative distribution. Logarithmic evaluations of the cumulative-probability difference avoid loss of precision in the tails. Conditional filling-factor draws follow a normal distribution with mean $B/A$ and variance $1/A$, truncated to $[0,1]$.

We integrate the posterior with cell-volume weights in the three logarithmic coordinates. Subdividing the original grid intervals by factors of two and four gives $77\times49\times49$ and $153\times97\times97$ integration positions, respectively; intrinsic spectra are interpolated rather than recomputed at every position. Nonfinite or nonpositive spectra are excluded. The one-component log evidence changes by less than 0.02 between these refinements (Table~\ref{tab:appendix_co_validation}).

Reported intervals are the marginalized 16th--84th percentiles. Derived quantities such as $n_{\rm H_2}T_{\rm kin}$ are calculated from joint samples. Pointwise median line predictions agree with every measured intensity to within $1.3\sigma_i$; their intervals represent parameter uncertainty, without added measurement noise.

For the additive two-component model (Equation~\ref{eq:two_component_model_method}), both components use the priors in Table~\ref{tab:radex_grid}, with independent filling factors and $T_1<T_2$. Nested sampling uses 400 live points, a stopping tolerance of $\Delta\ln Z=0.1$, and interpolation in logarithmic physical coordinates; interpolation cells with invalid corners are rejected.

We define $\Delta\ln Z_{21}=\ln Z_{\rm two}-\ln Z_{\rm one}$ and $B_{21}=\exp(\Delta\ln Z_{21})$, where $Z$ is the likelihood integrated over the normalized parameter prior. Table~\ref{tab:appendix_co_validation} reports ranges from repeated calculations, rather than credible intervals. The resulting $B_{21}=0.96$--$2.26$ gives no compelling preference for a second CO component under these priors.

\begin{table}[!h]
\centering
\caption{CO-only fit diagnostics and repeated model comparisons.}
\label{tab:appendix_co_validation}
\small\setlength{\tabcolsep}{3pt}
\begin{tabular}{lccccc}
\hline
Region & $\chi^2_{2}$ & $\chi^2_{4}$ & $\Delta\ln Z_{\rm grid}$ & $\Delta\ln Z_{21}$ & $B_{21}$ \\
\hline
R1 & $2.23$ & $2.23$ & $-0.005$ & $[0.60,0.66]$ & $[1.82,1.93]$ \\
R2 & $4.08$ & $4.02$ & $-0.018$ & $[-0.04,0.15]$ & $[0.96,1.16]$ \\
R3 & $2.74$ & $2.67$ & $+0.000$ & $[0.31,0.62]$ & $[1.36,1.86]$ \\
R4 & $1.19$ & $1.11$ & $+0.006$ & $[0.48,0.82]$ & $[1.62,2.26]$ \\
\hline
\end{tabular}
\tablecomments{Subscripts 2 and 4 denote grid-refinement factors, not component counts; $\Delta\ln Z_{\rm grid}=\ln Z_4-\ln Z_2$ compares one-component integrations. $\Delta\ln Z_{21}$ and $B_{21}$ span two repeated one-versus-two-component comparisons per region, with numerical errors of 0.19--0.21 on individual $\Delta\ln Z_{21}$ estimates.}
\end{table}

\subsection{Aperture-size sensitivity and direct-evaluation checks}
\label{app:aperture_sensitivity}

We repeat the CO fits at nine aperture diameters, $0.32$, $0.38$, $0.48$, $0.56$, $0.64$, $0.72$, $0.80$, $0.88$, and $0.96\arcsec$, using fractional-pixel extraction and the same regional linewidths and priors. All seven intensities enter the likelihood, including weak upper-transition measurements. The $0.96\arcsec$ endpoint is an internal control for this extraction series, distinct from the fiducial measurements.

Intrinsic intensities are linearly interpolated in logarithmic parameter coordinates. For each aperture, 1400 posterior proposal draws are checked with direct RADEX evaluations and importance weights
\begin{equation}
 w_k\propto
 \frac{\overline{\mathcal{L}}_{\rm direct}(\boldsymbol{\theta}_k)}
      {\overline{\mathcal{L}}_{\rm interp}(\boldsymbol{\theta}_k)},
 \qquad \sum_k w_k=1.
\end{equation}
The minimum effective sample size, $(\sum_k w_k^2)^{-1}$, is 1224 of 1400. At the $0.96\arcsec$ endpoints, the 95th-percentile absolute interpolation errors are $0.28$--$0.39\sigma_i$. Direct evaluations at one original grid position per region reproduce the stored spectra to fractional differences below $10^{-12}$.

Aperture-series contours use the interpolated grid; marginal intervals use the importance-corrected samples. Nested apertures share emission and calibration errors, so overlapping intervals are descriptive comparisons, not independent constraints on a common component.

\section{Supplementary nuclear comparison}
\label{app:izumi_comparison}

We test 18 illustrative conditions motivated by \citet{Izumi_2020}: $T_{\rm kin}=300$ and 500~K at $n_{\rm H_2}=10^4$ and $10^5\,\mathrm{cm^{-3}}$, and 100 and 200~K at $10^5\,\mathrm{cm^{-3}}$, each with $Q=3\times10^{16}$, $10^{17}$, and $3\times10^{17}\,\mathrm{cm^{-2}\,(km\,s^{-1})^{-1}}$. We interpolate the CO grid with the uniform-sphere geometry, $T_{\rm bg}=2.73$~K, and $\Delta v=190~\mathrm{km\,s^{-1}}$, fitting a single $0\leq\phi\leq1$ to all seven transitions.

Extractions are centered at $23^{\rm h}03^{\rm m}15.617^{\rm s}$, $+08^{\circ}52'26.00''$ (ICRS): a $0.38\arcsec$-diameter circle at the native $0.32\arcsec$ beam, and a central measurement after smoothing to a $0.38\arcsec$ Gaussian beam. Both use signed, baseline-subtracted intensities over $[-250,+250]~\mathrm{km\,s^{-1}}$, with channel-noise and calibration errors, no absorption correction, and the fixed nuclear linewidth. These extraction conventions differ from the final ratio maps and flux-fraction measurements.

The minimum seven-line $\chi^2$ values are 59.7 and 53.3 for the circle and smoothed-beam measurements, respectively. These selected homogeneous descriptions do not reproduce the full observed CO SLED. The test extends selected conditions to additional transitions; it neither samples the \citet{Izumi_2020} posterior nor excludes all conditions compatible with their multi-tracer analysis.

\section{Exploratory CO--H$_2$ comparison}
\label{app:h2_comparison}

We use the extinction-corrected H$_2$ S(1)--S(7) fluxes and quoted errors of \citet{Lai_2022}, with separate CO extractions in nominal $0.8\arcsec$ circles at their centers. R2--R4 correspond to LR7, LR9, and LR1, respectively. Point-spread functions, aperture corrections, and registration are not jointly matched. The H$_2$ measurements are from the literature; the excitation diagnostics and predictions below are calculated here.

Table~\ref{tab:h2_compact} gives apparent rotational temperatures from successive ortho-H$_2$ line pairs. Same-spin pairs avoid an assumed emitting ortho-to-para ratio (OPR). Quoted flux errors are propagated independently; extinction and calibration covariance are unavailable. These rotational temperatures can differ from kinetic temperatures under non-LTE excitation and are not imposed as CO-model priors.

\begin{table}[!h]
\centering
\caption{Apparent H$_2$ rotational temperatures.}
\label{tab:h2_compact}
\small\setlength{\tabcolsep}{5pt}
\begin{tabular}{lccc}
\hline\hline
Region & S(1)/S(3) & S(3)/S(5) & S(5)/S(7) \\
 & \multicolumn{3}{c}{(K)} \\
\hline
R2 & $358\pm5$ & $737\pm12$ & $1100\pm52$ \\
R3 & $436\pm42$ & $761\pm19$ & $1052\pm31$ \\
R4 & $353\pm7$ & $776\pm16$ & $1036\pm21$ \\
\hline
\end{tabular}
\tablecomments{Calculated from the extinction-corrected fluxes and quoted errors of \citet{Lai_2022} in the nominal $0.8\arcsec$ comparison apertures. Uncertainties are rounded upward to the nearest kelvin.}
\end{table}

H$_2$ predictions use the optically thin, collisionally excited non-LTE populations of \citet{Neufeld_2026}, version 2 \citep{Neufeld_grid_2026}, spanning $100$--$3981$~K and $10^1$--$10^{12}\,\mathrm{cm^{-3}}$. The collider OPR is fixed at 3; the emitting OPR varies independently. Populations, normalized separately within each spin isomer, are interpolated linearly on logarithmic temperature and density coordinates. Below 100~K we use an LTE screening approximation. Atomic and electron collisions, UV pumping, and H$_2$ radiative transfer are omitted.

For component $c$, the molecular column is $N_{{\rm H}_2,c}=Q_c\Delta v/x_{\rm CO}$, where $x_{\rm CO}=N_{\rm CO}/N_{\rm H_2}$. If $f_{u,c}$ is the upper-state population fraction within its spin isomer and $s_c$ is the fraction in that isomer, the model flux is
\begin{equation}
 F_{ul}=\frac{A_{\rm ap}}{4\pi D_L^2} A_{ul}h\nu_{ul}
 \sum_c\phi_c N_{{\rm H}_2,c}s_c f_{u,c}.
 \label{eq:app_h2_flux}
\end{equation}
Here $A_{\rm ap}$ is the projected aperture area, evaluated with $330$~pc per arcsecond, and $D_L=70.6$~Mpc. The spin fractions are $s_c=\mathrm{OPR}_c/(1+\mathrm{OPR}_c)$ for ortho-H$_2$ and $1/(1+\mathrm{OPR}_c)$ for para-H$_2$. We explore $x_{\rm CO}\in[0.5,2]\times10^{-4}$, shared by the two components in each aperture, and independent emitting OPRs in $[0.1,3]$.

We tested both the quoted independent H$_2$ errors and an illustrative fractional allowance,
\begin{equation}
 \sigma_{F,i,\epsilon}^2=\sigma_{F,i,\rm quoted}^2+(\epsilon F_i)^2,
 \qquad \epsilon=0.30.
\end{equation}
The 30\% allowance is a sensitivity test, not a measured calibration uncertainty or a covariance model; it is separate from the adopted CO errors.

A single-temperature H$_2$-only grid search, with variable density and separate spin columns, gives minimum $\chi^2=642$, 413, and 729 in R2--R4 using quoted errors, or 17.3, 11.5, and 17.0 with the allowance. These minima reach the high-density grid boundary. The least-discrepant sampled one-component CO descriptions leave H$_2$ $\chi^2\simeq52$--$53$ even with the allowance. The discrepancy therefore persists for the tested homogeneous descriptions under these excitation assumptions.

Reweighting the CO-only two-component samples for the comparison apertures and profiling over abundance and emitting OPRs leaves effective sample sizes of approximately 6, 3, and 1. We therefore report optimized illustrative examples, without reporting a joint posterior or joint Bayes factors.

\subsection{Illustrative two-component solutions}
\label{app:joint_examples}

\begin{figure}[!h]
\centering
\includegraphics[width=\textwidth]{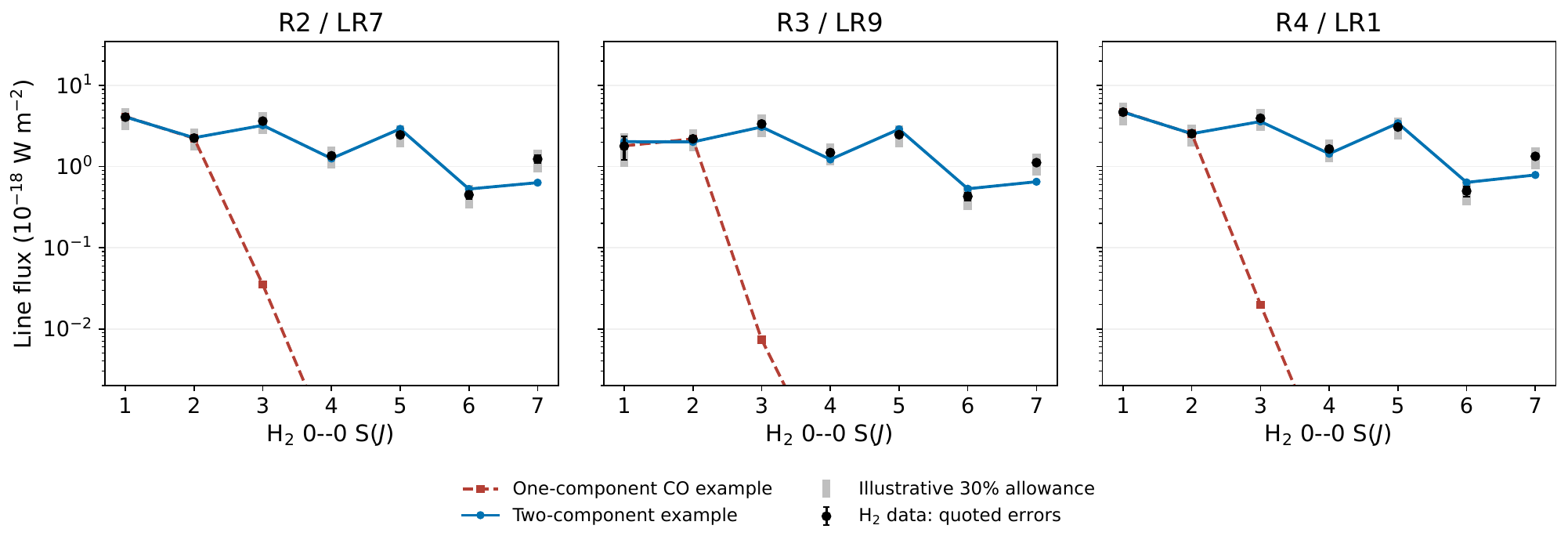}
\caption{Published extinction-corrected H$_2$ fluxes in the nominal $0.8\arcsec$ R2/LR7, R3/LR9, and R4/LR1 apertures \citep{Lai_2022}. Black error bars show quoted uncertainties; gray bars include the illustrative 30\% allowance. Red dashed curves are our least-discrepant sampled one-component CO predictions, and blue curves are optimized two-component CO--H$_2$ examples (Tables~\ref{tab:appendix_joint_parameters} and~\ref{tab:appendix_joint_diagnostics}). Some red predictions lie below the plotting range.}
\label{fig:h2_component_comparison}
\end{figure}

Multistart optimization supplies the examples in Table~\ref{tab:appendix_joint_parameters} and Figure~\ref{fig:h2_component_comparison}, using the 30\% allowance and $Q\leq10^{18.5}\,\mathrm{cm^{-2}\,(km\,s^{-1})^{-1}}$. Maximum checked H$_2$ optical depths are $2\times10^{-4}$, 0.071, and 0.002 in R2--R4, respectively, consistent with the optically thin approximation. CO predictions are checked by direct RADEX evaluation.

\begin{table}[!h]
\centering
\caption{Parameters of illustrative two-component CO--H$_2$ solutions in the ring.}
\label{tab:appendix_joint_parameters}
\small\setlength{\tabcolsep}{4pt}
\begin{tabular}{llrrrrrr}
\hline
Region & Component & $T$ & $\log n$ & $\log Q$ & $\phi$ & OPR & $10^4x_{\rm CO}$ \\
 & & (K) & & & & & \\
\hline
R2 & Cooler & $243$ & $3.52$ & $16.64$ & $0.1977$ & $0.57$ & $2.00$ \\
R2 & Hotter & $827$ & $7.00$ & $15.11$ & $0.0435$ & $3.00$ & $2.00$ \\
R3 & Cooler & $188$ & $3.53$ & $16.72$ & $0.3197$ & $0.10$ & $0.58$ \\
R3 & Hotter & $836$ & $7.00$ & $18.50$ & $6.57\times10^{-6}$ & $3.00$ & $0.58$ \\
R4 & Cooler & $180$ & $3.58$ & $17.07$ & $0.2053$ & $0.17$ & $0.50$ \\
R4 & Hotter & $841$ & $7.00$ & $15.00$ & $0.0176$ & $3.00$ & $0.50$ \\
\hline
\end{tabular}
\tablecomments{Optimized examples for nominal $0.8\arcsec$ apertures, using the 30\% H$_2$ allowance. Here $n=n_{\rm H_2}$ is in $\mathrm{cm^{-3}}$, $Q$ in $\mathrm{cm^{-2}\,(km\,s^{-1})^{-1}}$, and $x_{\rm CO}=N_{\rm CO}/N_{\rm H_2}$ is shared within each region. OPR is the emitting H$_2$ ratio. Digits identify the examples, not parameter precision.}
\end{table}

\begin{table}[!h]
\centering
\caption{Fit diagnostics and tracer contributions of the illustrative ring solutions.}
\label{tab:appendix_joint_diagnostics}
\small\setlength{\tabcolsep}{4pt}
\begin{tabular}{lrrrrrrr}
\hline
Region & $\chi^2_{\rm CO}$ & $\chi^2_{\rm H_2}$ & $\chi^2_{\rm H_2,0}$ & $\chi^2_{\rm H_2,OPR3}$ & $\epsilon_{\rm CO,max}$ & $f_{M,h}$ & $f_{8,h}$ \\
 & & & & & ($\sigma$) & (\%) & (\%) \\
\hline
R2 & $7.85$ & $3.26$ & $42.57$ & $8.31$ & $0.30$ & $0.65$ & $9.85$ \\
R3 & $6.34$ & $3.31$ & $78.94$ & $9.25$ & $0.24$ & $0.12$ & $0.58$ \\
R4 & $1.35$ & $2.94$ & $97.92$ & $8.54$ & $0.33$ & $0.07$ & $1.50$ \\
\hline
\end{tabular}
\tablecomments{$\chi^2_{\rm CO}$ (direct RADEX) and $\chi^2_{\rm H_2}$ describe Table~\ref{tab:appendix_joint_parameters}. The next two columns are separate optimizations: quoted H$_2$ errors alone ($0$), or fixed emitting OPRs of 3 with the allowance (OPR3). $\epsilon_{\rm CO,max}$ is the maximum absolute direct-minus-interpolated CO difference in observational-error units. $f_{M,h}$ and $f_{8,h}$ are the modeled hot-component mass and CO(8--7) fractions.}
\end{table}

The examples fit 14 line measurements with 11 adjustable parameters. All hotter-component densities reach $10^7\,\mathrm{cm^{-3}}$; the tabulated parameters identify optimized solutions, not posterior constraints. Table~\ref{tab:appendix_joint_diagnostics} gives their fit statistics and separate optimizations using quoted H$_2$ errors alone or fixed emitting OPRs of 3. Substantial residuals remain with quoted errors alone. The small CO contribution from the hotter component illustrates how an effective one-component CO description can coexist with additional H$_2$ emission; the CO-only evidence does not establish this decomposition.

For common abundance and projected aperture area, the example hot-component molecular-mass fraction is
\begin{equation}
 f_{M,h}=\frac{\phi_h Q_h\Delta v_h}
 {\sum_c\phi_c Q_c\Delta v_c},
\end{equation}
and its CO-line fraction is
\begin{equation}
 f_{J,h}=\frac{\phi_h m_{J,h}}
 {\sum_c\phi_c m_{J,c}}.
\end{equation}
The fractions in Table~\ref{tab:appendix_joint_diagnostics} depend on the common adopted abundance and the optimized component columns and filling factors; they are illustrative tracer contributions, not measured mass partitions.


\section{Aperture measurements}
\label{app:continuum_measurements}
\begin{table}[!h]
\centering
\caption{Continuum flux densities in the fiducial $0.96\arcsec$-diameter apertures at the listed frequencies (GHz).}
\label{tab:continuum_fluxes}
\small\setlength{\tabcolsep}{3pt}
\begin{tabular}{lcccc}
\hline\hline
Region & 461.041 & 691.473 & 806.652 & 921.800 \\
 & \multicolumn{4}{c}{(mJy)} \\
\hline
R1 & $24.52\pm4.91$ & $97.66\pm19.54$ & $176.14\pm35.25$ & $202.87\pm40.60$ \\
R2 & $8.81\pm1.77$ & $36.37\pm7.29$ & $58.08\pm11.68$ & $91.26\pm18.31$ \\
R3 & $9.65\pm1.94$ & $38.79\pm7.77$ & $63.91\pm12.84$ & $110.48\pm22.14$ \\
R4 & $14.20\pm2.85$ & $52.42\pm10.49$ & $89.84\pm18.01$ & $131.80\pm26.40$ \\
\hline
\end{tabular}
\tablecomments{These are the measured inputs to the four-frequency modified-blackbody fits in Section~\ref{sec:continuum_method}. The supplied fit treats their uncertainties as independent; interband calibration covariance is not included. Fluxes are rounded to two decimals and uncertainties upward to the same precision.}
\end{table}

\begin{table}
\centering
\caption{CO integrated brightness temperatures in the
fiducial apertures.}
\label{tab:aperture_co_intensities}
\small
\setlength{\tabcolsep}{4pt}
\begin{tabular}{lcccc}
\hline\hline
Transition & R1 & R2 & R3 & R4 \\
 & \multicolumn{4}{c}{($\mathrm{K\,km\,s^{-1}}$)} \\
\hline
CO(1--0) & $987\pm53$   & $470\pm33$ & $542\pm31$ & $753\pm42$ \\
CO(2--1) & $1842\pm185$ & $793\pm80$ & $779\pm78$ & $931\pm94$ \\
CO(3--2) & $1884\pm189$ & $725\pm73$ & $705\pm71$ & $836\pm84$ \\
CO(4--3) & $1756\pm176$ & $531\pm54$ & $534\pm54$ & $641\pm65$ \\
CO(6--5) & $954\pm191$  & $271\pm55$ & $282\pm57$ & $324\pm66$ \\
CO(7--6) & $407\pm83$   & $102\pm26$ & $191\pm40$ & $238\pm52$ \\
CO(8--7) & $255\pm53$   & $109\pm24$ & $47\pm16$  & $97\pm22$ \\
\hline
\end{tabular}
\tablecomments{
Measurements use the $0.96\arcsec$-diameter apertures
adopted for the fiducial CO modeling.
Uncertainties combine statistical errors and calibration
contributions of 5\% for CO(1--0), 10\% for CO(2--1),
CO(3--2), and CO(4--3), and 20\% for CO(6--5),
CO(7--6), and CO(8--7).
The complete hexagonal-grid measurements are provided
in the machine-readable tables accompanying this article.
}
\end{table}

\clearpage
\bibliography{reference.bib}

\end{document}